\documentclass[12pt]{article}
\usepackage{natbib,amssymb,times,amsthm,amsmath, macros_genop, booktabs}
\usepackage[pdftex]{graphicx}
\renewcommand{\baselinestretch}{1.85}

\usepackage{xcolor}

\usepackage[letterpaper, left=1in, top=1in, right=0.75in, bottom=1in, nohead,verbose, ignoremp]{geometry}

\graphicspath{{./}}

\begin{document}

\title{ \vspace{-2.5cm} Bayesian Elastic Net Regression \\\vspace{-0.5cm} with Structured Prior Dependence}

\author{Christopher M.~Hans\footnote{Corresponding author: hans@stat.osu.edu}~and Ningyi Liu\\\vspace{-0.5cm}
{\small Department of Statistics, The Ohio State University, Columbus, OH, 43210, USA}}

\date{\small December 2025}
\maketitle

\vspace{-1.25cm} \abstract{
Many regularization priors for Bayesian regression
assume the regression coefficients are \emph{a priori} independent.  In particular this is the case
for standard Bayesian treatments of the lasso and the elastic net.  While independence may be reasonable 
in some data-analytic settings, incorporating dependence in these prior distributions provides 
greater modeling flexibility.  This paper introduces the orthant normal distribution in its general form
and shows how it can be used to structure prior dependence in the Bayesian elastic net regression model.
An $\ell_1$-regularized version of Zellner's $g$ prior is introduced as a special case, creating a new 
link between the literature on penalized optimization and an important class of regression priors.
Computation is challenging due to an intractable normalizing constant in the prior. We avoid this issue
by modifying slightly a standard prior of convenience for the hyperparameters in such a way to enable
simple and fast Gibbs sampling of the posterior distribution. The benefit of including structured prior
dependence in the Bayesian elastic net regression model is demonstrated through simulation and a
near-infrared spectroscopy data example.
}


\newpage

\section{Introduction}\label{sec:intro}
Bayesian regression models with connections to penalized optimization procedures
have received extensive attention in the literature.
Work on Bayesian models  \citep{park:08, hans:09, hans:10} related to the lasso 
\citep{tibs:96} led  to interest in other penalized optimization settings.  
\citet{li:10}, \citet{hans:11}, and \citet{roy:17} describe how the elastic net penalty 
function \citep{zou:05} can be used to generate a class of prior distributions on regression 
coefficients. Work along these lines dovetails with a related stream of work on Bayesian shrinkage 
priors. \citet{grif:10, grif:11}, \citet{arma:13}, and \citet{bhat:15} discuss priors that generalize those 
underlying the Bayesian lasso, and \citet{carv:10} and \citet{pols:11, pols:12, pols:12b}, who describe 
priors constructed to have particular shrinkage profiles.

One feature that is common to most of these prior distributions is that the regression 
coefficients are assumed to be conditionally independent given a set of hyperparameters. This feature
is also shared by many ``off-the-shelf'' prior distributions for regression coefficients,
e.g.~Bayesian treatments of ridge regression \citep{hoer:70}. 
While the conditional independence assumption requires little input from the user and
often results in straightforward computation, there are many situations where it
would be beneficial to allow for \emph{a priori} dependence among the 
regression coefficients.  If substantive prior information is available
(e.g., if it is thought that two coefficients should likely not have the same sign), a
good analysis would incorporate such information into the model.  \citet{grif:12}
describe correlated priors for regression coefficients where dependence is
introduced because of known structural features of the model (e.g., coefficients
related to factor or categorical variables, and first-order Markov dependence structures 
arising from data observed over time).  \citet{bedr:96} describe priors for regression
coefficients in generalized linear models where dependence is induced through elicitation
of prior information about the regression surface.  Other priors exhibiting (conditional)
\emph{a priori} dependence have been introduced due to computational or theoretical
considerations.  Zellner's $g$ prior \citep{zell:86} is a notable example.  Extensions of
the $g$ prior and additional motivation for this particular form of prior dependence can
be found in \cite{zell:80}, \citet{west:03}, \citet{cui:08}, \citet{lian:08}, \citet{maru:11}, 
\citet{baya:12}, \citet{maru:14} and \citet{som:14}.

The ability to incorporate prior dependence for regression
coefficients in Bayesian regularized regression---in particular for models inspired
by penalized optimization procedures like the lasso and elastic net---is desirable.
This paper describes one such way in which prior dependence can be incorporated
in these settings.  The approach involves generalizing the orthant normal prior distribution
\citep{hans:11} to allow for \emph{a priori} dependence among the regression coefficients.
A special case of the generalized orthant normal prior is shown to be an
$\ell_1$-regularized version of Zellner's $g$ prior, providing a new Bayesian link
between the literature on penalized optimization procedures and an important class
of regression priors.  Full Bayesian inference on all parameters---including prior
hyperparameters---is often computationally challenging for regression models with
prior dependence.  Such challenges arise for the generalized orthant normal prior,
and strategies for addressing these challenges are introduced. After providing an
illustration of posterior inference under the $\ell_1$-regularized $g$ prior, we
demonstrate the usefulness of including structured dependence in the Bayesian elastic 
net regression model via simulation and a near-infrared spectroscopy data example.

\section{Generalized orthant normal priors}\label{sec:genop}

\subsection{The orthant normal distribution}\label{sec:onorm}
We consider the normal linear model,
$y \mid \alpha, \beta, \sigma^2 \sim \mbox{N}(\alpha 1_n + X\beta, \sigma^2 I_n)$, where 
$y$ is an $n$-vector of observations and $X$ is an $n \times p$ matrix of predictors.
Throughout the paper we assume $n>p$ and focus on the role of the orthant normal distribution 
as a prior distribution for regression coefficients, $\beta$. In its most general form, we 
define the orthant normal distribution to have density function
\begin{equation}
	p(\beta \mid \omega, \mu, \Sigma) = \sum_{z \in \mathcal{Z}}
		\omega_z \mbox{N}^{[z]}(\beta \mid \mu_z, \Sigma_z).
		\label{eq:genop}
\end{equation}
In this notation, $\mathcal{Z} = \{-1, 1\}^p$ represents the collection of the $2^p$ 
\emph{orthants} of $\mathbb{R}^p$: for $p$-vectors $z \in \mathcal{Z}$ and $\beta \in \mathbb{R}^p$,
if $\beta_j \geq 0 \Leftrightarrow z_j = 1$ and $\beta_j < 0 \Leftrightarrow z_j = -1$, then
$\beta$ is said to lie in orthant $\mathcal{O}_z$.  The density in (\ref{eq:genop})
is a weighted sum of $2^p$ properly-normalized probability density functions for 
orthant-truncated normal distributions,
\[
	\mbox{N}^{[z]}(\beta \mid m, S) \equiv \frac{\mbox{N}(\beta \mid
		m, S)}{\mbox{P}(z, m, S)}1(\beta \in \mathcal{O}_z),
\]
where
\[
	\mbox{N}(\beta \mid m, S) = (2\pi)^{-p/2}|S|^{-1/2}
		\exp\{-(\beta - m)^T S^{-1}(\beta - m)/2\}
\]
is the density function for a multivariate normal distribution,
\[
	\mbox{P}(z, m, S) = \int_{\mathcal{O}_z} \mbox{N}(t \mid m, S)dt
\]
is the integral of a multivariate normal density function over a particular
orthant $\mathcal{O}_z$ of $\mathbb{R}^p$, and $1(\cdot)$ is a $0/1$ binary indicator 
function. The $2^p$ parameters 
$\omega_z \geq 0$ are orthant-specific weights that sum to one, and the
$\mu_z$ and $\Sigma_z$ are orthant-specific location vectors and positive-definite
dependence matrices for the orthant-truncated normal distributions.  We use the terms
``location'' and ``dependence'' rather than ``mean'' and ``covariance'' to avoid confusion
because these parameters are the means and covariances of the underlying normal distributions
but not of the \emph{truncated} normal distributions. The orthant normal prior distribution as 
described in (\ref{eq:genop}) is quite general: depending on how $\omega_z$, $\mu_z$ and 
$\Sigma_z$ are chosen, the density function may not be everywhere differentiable, or even 
continuous. Our interest lies in formulations where the density function is 
everywhere continuous but not necessarily everywhere differentiable.  

\subsection{Independence in orthant normal priors}\label{sec:iid}
\begin{figure}[ht]
 \begin{center}
 \includegraphics[scale=0.55]{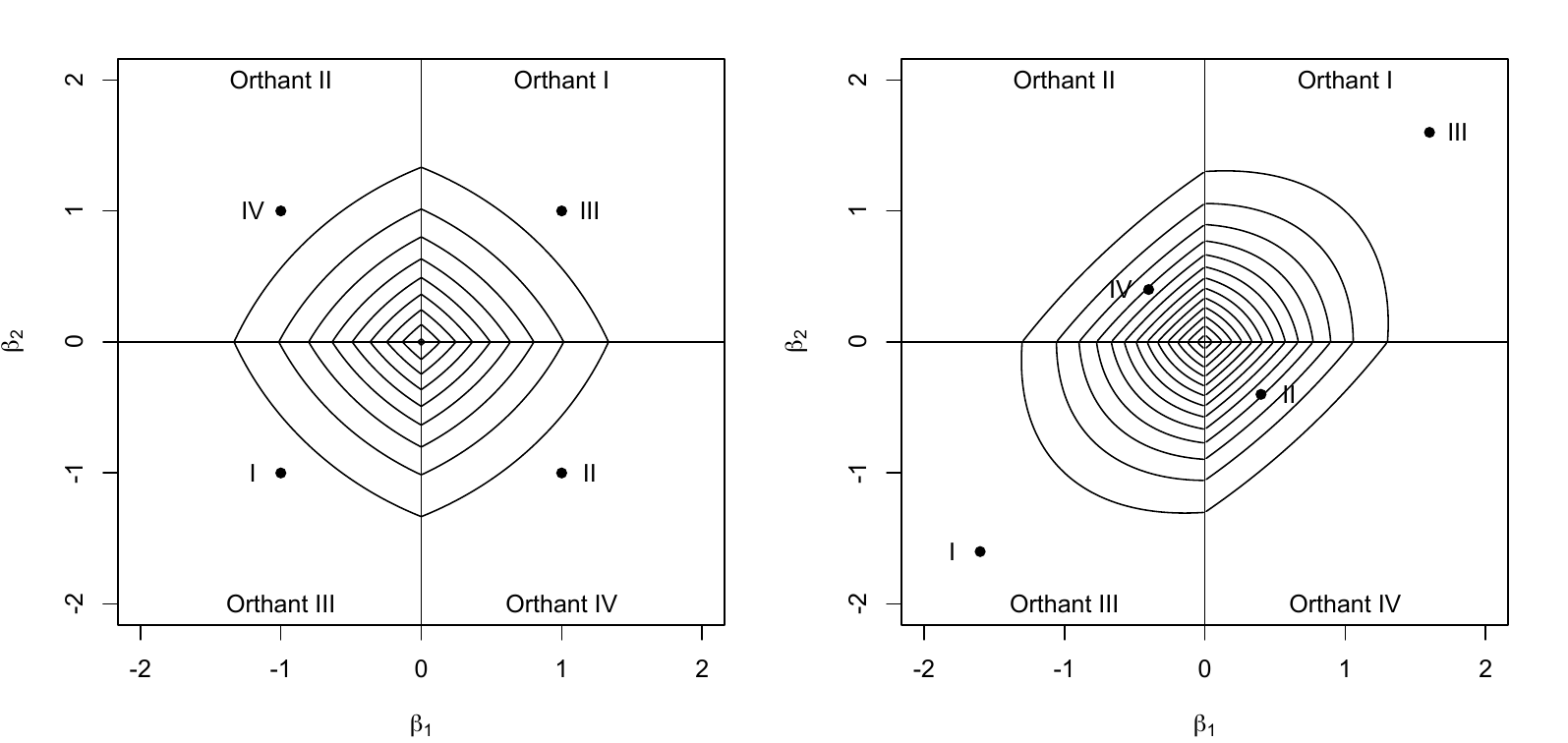}
 \caption{Contours of two orthant normal density functions when $p = 2$.  
 		The example in
 		the left panel corresponds to \emph{a priori} conditional
		independence of the regression coefficients given $\sigma^2$, while
		the right panel incorporates a conditional dependence structure.
		The points correspond to the orthant-specific location
		vectors $\mu_z$ (labeled by the orthant to which they belong).
 	        \label{fig:examples}}
 \end{center}
\end{figure}

The orthant normal distribution arises in regression modeling because,
when used as a prior distribution for regression coefficients,
particular special cases give rise to the
``Bayesian elastic net'' and ``Bayesian lasso'' models.
\citet{hans:11} introduced the special case where
\begin{equation}
	\omega_z = 2^{-p}, \;\; \mu_z = -\frac{\lambda_1}{2\lambda_2}z, 
	\;\;\mbox{   and   }\;\; \Sigma_z = \frac{\sigma^2}{\lambda_2} I_p,
	\label{eq:enet}
\end{equation}
and showed that
simplification of the log prior density function yields
\begin{equation}
	-2\sigma^2 \log p(\beta \mid \lambda_1, \lambda_2, \sigma^2) = \mbox{const.} +  
		\lambda_1 \sum_{j=1}^p |\beta_j| + \lambda_2 \sum_{j=1}^p \beta_j^2,
		\label{eq:enetpen}
\end{equation}
which is the elastic net penalty function \citep{zou:05} with penalty parameters $\lambda_1 > 0$ 
and $\lambda_2 > 0$.  The limiting case $\lambda_2 \rightarrow 0$ corresponds to the lasso 
penalty function.  The left panel of Figure~\ref{fig:examples} displays this prior density function
for a two-dimensional example with $\lambda_1 = 2$, $\lambda_2 = 1$ and $\sigma^2 = 1$.
The density function is everywhere continuous but is not differentiable along the
coordinate axes.  The points labeled on the plot are the four orthant-specific location vectors, e.g.,
the point labeled ``II'' that lies in the fourth orthant (quadrant) represents
the $\mu_z$ vector that corresponds to the location vector for orthant II where $z = (-1, 1)^T$.

\renewcommand{\baselinestretch}{1}
\renewcommand{\baselinestretch}{1.85}

Conditional independence of the $\beta_j$ given $\sigma^2$, $\lambda_1$, and $\lambda_2$ under 
the ``elastic net'' special case 
of the orthant normal prior can be seen by noting that the prior density under (\ref{eq:enet}) can also 
be written as
\[
	p(\beta \mid \sigma^2, \lambda_1, \lambda_2) = \prod_{j=1}^p
		\left\{ (0.5)\cdot \mbox{N}^{-}\left( \beta_j \left| \frac{\lambda_1}{2\lambda_2},
		\frac{\sigma^2}{\lambda_2}\right)\right.
		+
		(0.5)\cdot \mbox{N}^{+}\left( \beta_j \left| -\frac{\lambda_1}{2\lambda_2},
		\frac{\sigma^2}{\lambda_2}\right)\right.
		\right\},
\]
where
\[
	\mbox{N}^-(x \mid a, b^2) = \frac{\mbox{N}(x \mid a, b^2)1(x < 0)}
		{\Phi(-a/b)}
		\;\;\;\mbox{ and }\;\;\;
	\mbox{N}^+(x \mid a, b^2) = \frac{\mbox{N}(x \mid a, b^2)1(x \geq 0)}
		{\Phi(a/b)}
\]
are the density functions for $\mbox{N}(a,b^2)$ random variables truncated to be negative and
non-negative, respectively, with $\Phi$ representing the standard normal cdf.
This conditional \emph{a priori} i.i.d.~assumption is shared by many of the 
regularization priors for Bayesian regression surveyed in Section~\ref{sec:intro}.
While this may be reasonable in some settings, having the ability to directly incorporate 
dependence in the prior distribution would allow for more flexibility in constructing
models and, ultimately, estimators and predictions.  The purpose of this paper is to describe a 
straightforward mechanism for incorporating prior dependence in such models using the orthant 
normal prior.

\subsection{Dependence in orthant normal priors}\label{sec:dep}
The simplest means of incorporating prior dependence in the orthant normal prior is
to relax the assumption that $\Sigma_z \propto I_p$. We must take care when relaxing this assumption, 
though, because without any restrictions on $\Sigma_z$, 
$\omega_z$, and $\mu_z$ we allow for priors whose density functions may not be 
everywhere continuous.  Leaving open the possibility that this may be desirable in certain situations, 
we will exclude this case from the collection of priors described in this paper.
We generalize the i.i.d.~orthant normal prior defined by (\ref{eq:enet}) to incorporate prior 
dependence while guaranteeing continuity of the prior density function by
specifying
\begin{equation}
	\mu_z = -\frac{\lambda_1}{2\lambda_2}\Sigma z, 
	\;\; \Sigma_z = \frac{\sigma^2}{\lambda_2} \Sigma,
	\;\;\mbox{   and   }\;\; \omega_z = \omega^{-1} \frac{\mbox{P}(z, \mu_z, \Sigma_z)}
		{\mbox{N}(0 \mid \mu_z, \Sigma_z)},
	\label{eq:genen}
\end{equation}
where
\[
	\omega \equiv \sum_{z \in \mathcal{Z}}  \frac{\mbox{P}(z, \mu_z, \Sigma_z)}
		{\mbox{N}(0 \mid \mu_z, \Sigma_z)}
\]
and $\Sigma$ is a positive-definite dependence matrix that is common to all
orthants.  
The orthant-specific location parameters in (\ref{eq:enet}) have
all been transformed by pre-multiplying by $\Sigma$ in (\ref{eq:genen}), and the
orthant weights are specified in (\ref{eq:genen}) to preserve continuity of the density function
(see Appendix~\ref{app:densprop} for details).
We believe this formulation of the prior strikes a good balance between simplicity---one need only
specify $\lambda_1$, $\lambda_2$ and a single matrix $\Sigma$---and 
flexibility.

\begin{figure}[!ht]
 \begin{center}
 \includegraphics[scale=0.5]{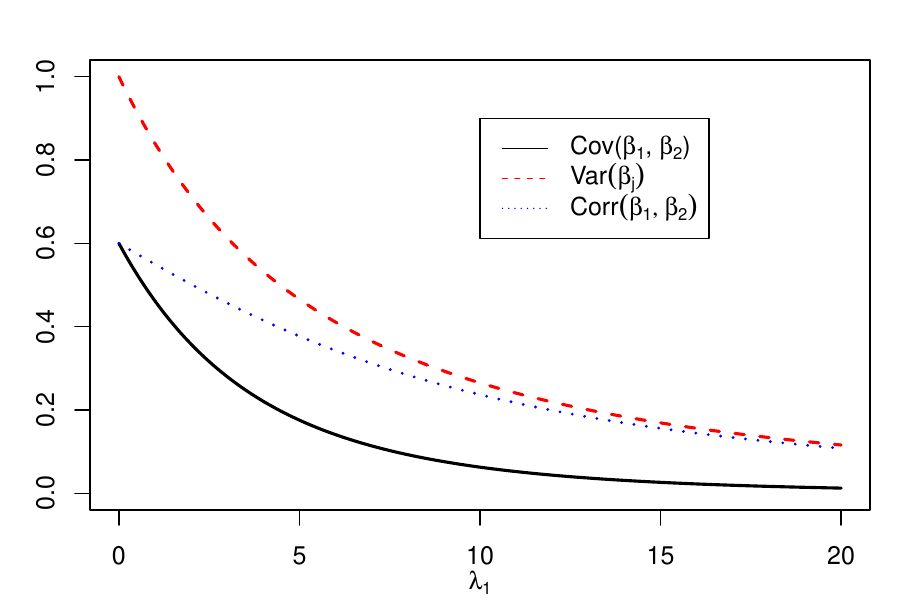}
 \caption{Aspects of the prior in the example in Section~\ref{sec:dep} when
 	$p=2$ as a function of $\lambda_1$.
 	\label{fig:lam1}}
 \end{center}
\end{figure}

The right panel of Figure~\ref{fig:examples} displays contours of the prior density function
for a two-dimensional example when $\lambda_1 = 6$, $\lambda_2 = 3$, $\sigma^2 = 3$,
$\Sigma_{11} = \Sigma_{22} = 1$ and $\Sigma_{12} = \Sigma_{21} = 0.6$.  The prior dependence
is clear, 
and the key features of the original Bayesian elastic net prior---within-orthant normality and 
non-differentiability along the coordinate axes---are preserved. While the orthant constraints 
give rise to $\ell_1$-like penalization, they also make interpreting the prior ``dependence'' 
matrix somewhat difficult, as $\sigma^2\lambda_2^{-1}\Sigma$ is not the prior covariance of 
$\beta$ (unless $\lambda_1 = 0$).  In this example, the prior covariance matrix is approximately
\[
	\mbox{Cov}(\beta) \approx \left( \begin{array}{ll} 0.41 & 0.14 \\ 0.14 & 0.41 \end{array}\right),
\]
which implies a prior correlation of approximately $0.34$.  The prior covariance depends
on all three parameters.  Figure~\ref{fig:lam1} shows how the prior dependence
changes as a function of $\lambda_1$ in this example for fixed $\sigma^2$, $\Sigma$ and $\lambda_2$.

\renewcommand{\baselinestretch}{1}
\renewcommand{\baselinestretch}{1.85}

\subsection{Orthant normal posteriors}\label{sec:onpost}
We focus for now on the posterior of $\beta$ given $\sigma^2$, $\lambda_1$, and $\lambda_2$
having integrated the intercept $\alpha$ out of the likelihood function under the usual
non-informative prior $p(\alpha) \propto 1$:
\begin{eqnarray}
	p(y \mid \beta, \sigma^2) &\equiv& \int_{-\infty}^\infty p(y \mid \alpha, \beta, \sigma^2)
		p(\alpha) d\alpha \nonumber \\
	&\propto& (2\pi\sigma^2)^{-(n-1)/2}n^{-1/2}e^{-\frac{1}{2\sigma^2}(y_c - X_c \beta)^T
		(y_c - X_c\beta)},
		\label{eq:intlik}
\end{eqnarray}
where $y_c$ and $X_c$ denote that $y$ and the columns of $X$ have been centered to have
mean zero. We use this likelihood function and the centered data throughout, dropping the ``$c$''
subscript in the notation.

Under the prior specification in (\ref{eq:genen}) and the likelihood defined in (\ref{eq:intlik}), 
it can be shown that the posterior distribution of $\beta$ given the observed data $y$ and the 
parameters $\sigma^2$, $\lambda_1$ and $\lambda_2$ is also an orthant normal distribution with 
updated orthant-specific location and structure parameters
\begin{equation}
	\mu_z^* = (\lambda_2 \Sigma^{-1} + X^TX)^{-1} \left(X^T y - \frac{\lambda_1}{2}z\right) 
	\label{eq:genpostmu}
\end{equation}
and
\begin{equation}
	\Sigma^*_z = \sigma^2 (\lambda_2 \Sigma^{-1} + X^TX)^{-1},
	\label{eq:genpostsig}
\end{equation}
the latter of which does not depend on $z$. The updated
orthant-specific weights $\omega^*_z$ are calculated as in (\ref{eq:genen}) using the updated 
parameters $\mu_z^*$ and $\Sigma_z^*$. Letting $G_{\lambda_2} = (\lambda_2 \Sigma^{-1} + 
X^TX)^{-1}$ and $\hat{\beta}_{G_{\lambda_2}} = G X^T y$, the following alternate expressions provide 
additional insight into the orthant-specific parameters:
\begin{eqnarray*}
	\mu_z^* &=& \hat{\beta}_{G_{\lambda_2}} - \frac{\lambda_1}{2}G z, \\
	\Sigma_z^* &=& \sigma^2 G_{\lambda_2}.
\end{eqnarray*}
The parameter $\hat{\beta}_{G_{\lambda_2}}$ would be the ridge regression estimate of $\beta$ if
$\Sigma = I_p$. Viewing $\hat{\beta}_{G_{\lambda_2}}$ as a generalized ridge regression 
estimate for a generic $\Sigma$, $\mu^*_z$ can therefore be viewed as an $\ell_1$-penalized
generalized ridge regression estimate.

\subsection{Connections to penalized optimization}\label{sec:penop}
Just as the orthant normal prior (\ref{eq:enet}) with $\Sigma = I_p$ has a connection
to the elastic net penalty function (\ref{eq:enetpen}), the generalized orthant normal prior
(\ref{eq:genen}) has a connection to a penalty function for a penalized optimization procedure.
As shown in Appendix~\ref{app:densprop}, the log prior density of the generalized orthant normal 
prior (\ref{eq:genen}) satisfies
\begin{equation}
	-2\sigma^2 \log p(\beta \mid \sigma^2, \lambda_1, \lambda_2) = 
		\mbox{const.} + \lambda_1 | \beta |_1+ 
		\lambda_2 \beta^T \Sigma^{-1} \beta,
		\label{eq:genpen}
\end{equation}
where $| \beta |_1 = \sum_{j=1}^p |\beta_j|$ is the $\ell_1$-norm of $\beta$.  This is equivalent
to the penalty function $\Omega(\beta) = \lambda_1 |\beta |_1 + \lambda_2 \beta^T \Lambda \beta$ (with
$\lambda_1 > 0$, $\lambda_2 > 0$ and $\Lambda$ a positive semidefinite matrix)
used by \cite{slaw:10} to define the ``structured elastic net.''  Including the matrix $\Lambda$ 
in the penalty function ``aims at capturing the a priori association structure (if available) of the features,''
and the authors discuss choices that correspond to different structural knowledge, e.g., features that
are sampled in time or space, or features that are connected via a graphical structure. Such features
can be incorporated analogously into the generalized orthant normal prior.  \citet{slaw:12}
extends this penalty function for use in quantile regression and support vector classification.

\subsection{$\ell_1$-regularized $g$ priors}\label{sec:gpriors}
An interesting special case of the generalized orthant normal prior relates to Zellner's $g$ prior
\citep{zell:86}.
The traditional $g$ prior
has the form
\begin{equation}
	\beta \mid \sigma^2, g \sim \mbox{N}\left(0, g \sigma^2 (X^T X)^{-1}\right),
	\label{eq:gprior}
\end{equation}
where $g > 0$ is a hyper-parameter that can be fixed \emph{a priori} or 
modeled as a random variable.
This prior features prominently in the contemporary
Bayesian variable selection literature
\citep[e.g.][]{west:03, cui:08, lian:08, maru:11, baya:12, som:14}.

The same within-orthant dependence structure as the $g$ prior can be obtained by
the generalized orthant normal prior by
setting $\Sigma = (X^T X)^{-1}$, yielding
\begin{equation}
	\mu_z = -\frac{\lambda_1}{2\lambda_2} (X^TX)^{-1}z \;\;\;
	\mbox{ and } \;\;\;
	\Sigma_z = \frac{\sigma^2}{\lambda_2}(X^TX)^{-1}.
	\label{eq:l1g}
\end{equation}
Setting $\lambda_1 = 0$ recovers Zellner's $g$ prior with
$g = \lambda_2^{-1}$.  Following (\ref{eq:genpostmu}) and (\ref{eq:genpostsig}), the 
conditional posterior $p(\beta \mid y, \sigma^2, \lambda_1, \lambda_2)$
is a generalized orthant normal distribution with
\begin{eqnarray}
	\mu_z^* &=& \frac{1}{1+\lambda_2}\left(\bols - \frac{\lambda_1}{2}(X^TX)^{-1} 
		z\right)  \label{eq:mupost} \\
	&=& \frac{1}{1 + \lambda_2}\bols + \frac{\lambda_2}{1+\lambda_2}\mu_z,
		\nonumber
\end{eqnarray}
where $\bols$ is the ordinary least squares estimate $(X^T X)^{-1} X^T y$, and
\begin{eqnarray}
	\Sigma_z^* &=& \frac{\sigma^2}{1+ \lambda_2}(X^TX)^{-1} \label{eq:sigpost} \\
	&=& \frac{\lambda_2}{1+\lambda_2} \Sigma_z.
	\nonumber
\end{eqnarray}
The updated orthant-specific
weights $\omega^*_z$ are calculated as in (\ref{eq:genen}) using the updated parameters
$\mu_z^*$ and $\Sigma_z^*$.
The conditional posterior distribution under Zellner's $g$ parameterized with $g = \lambda_2^{-1}$ 
is
\begin{equation}
	\beta \mid y, \sigma^2, \lambda_2 \sim \mbox{N}\left(
	\frac{1}{1 + \lambda_2}\bols, \frac{\sigma^2}{1+\lambda_2}(X^TX)^{-1}\right).
	\label{eq:gposterior}
\end{equation}
Comparing (\ref{eq:mupost}) and (\ref{eq:sigpost}) with the mean and covariance of (\ref{eq:gposterior}),
we can interpret the model that generates (\ref{eq:mupost}) and (\ref{eq:sigpost}) as an 
$\ell_1$-regularized version of the $g$ prior.

\subsection{Alternate parameterizations and representations}\label{sec:altparrep}
Several parameterizations of the Bayesian elastic net prior distribution when $\Sigma \propto I_p$ 
have appeared in the literature. \citet{li:10} and \citet{hans:11} parameterize the prior as
\[
	p(\beta \mid \sigma^2, \lambda_1, \lambda_2) \propto \exp\left\{-\frac{\lambda_2}{2\sigma^2}
		\beta^T\beta - \frac{\lambda_1}{2\sigma^2}|\beta|_1|\right\},
\]
where $\beta^T\beta$ and $|\beta|_1$ are scaled commonly by $2\sigma^2$. \citet{kyun:10}
and \citet{roy:17} scale these terms differentially:
\[
	p(\beta \mid \sigma^2, \lambda_1^*, \lambda_2) \propto
		\exp\left\{-\frac{\lambda_2}{2\sigma^2}\beta^T\beta -
		\frac{\lambda_1^*}{\sigma}|\beta|_1\right\}.
\]
This parameterization has the property that the units of $\lambda_1^*$ and $\lambda_2$
are free of the units of $y$. The generalized orthant normal prior under the differential 
scaling would have a density function that satisfies
\begin{equation}
	-2\sigma^2 \log p(\beta \mid \sigma^2, \lambda_1^*, \lambda_2) = \mbox{const.} +
		\lambda_2 \beta^T \Sigma^{-1}\beta	+ 2\sigma \lambda_1^* |\beta|_1
		\label{eq:diffpen}
\end{equation}
instead of (\ref{eq:genpen}). We use the commonly scaled parameterization satisfying
(\ref{eq:genpen}) throughout, but the model and all associated computation can be easily
modified to accommodate the differential scaling in (\ref{eq:diffpen}). We describe these
modifications briefly in Appendix~\ref{app:diffpar}.

In addition to this alternate parameterization, there are also alternate representations of the
Bayesian elastic net prior when $\Sigma \propto I_p$. \citet{li:10} and \citet{hans:11}
describe how the density function $p(\beta_j \mid \sigma^2, \lambda_1, \lambda_2)
\propto \exp\{-(\lambda_2 \beta_j^2 + \lambda_1 |\beta_j|)/(2\sigma^2)\}$ can be
represented as a scale mixture of normal distributions \citep{andr:74, west:87} by
introducing iid latent variables, $\tau_j$, $j=1, \ldots, p$. \citet{roy:17} does the same under 
the differentially scaled parameterization. This representation facilitates posterior sampling:
when $\Sigma \propto I_p$, the vector of regression coefficients, $\beta$, can be easily sampled 
from its full conditional distribution given the latent variables, $\{\tau_j\}$, and the latent 
variables can be sampled \emph{independently} from their respective full conditional distributions 
given $\beta$.  We note that such a representation for the orthant normal prior for general $\Sigma$
would not necessarily provide any computational advantages because the latent 
variables would not necessarily be conditionally independent or follow a distribution that
could be easily sampled \emph{a posteriori}. We work directly with the prior as represented
in (\ref{eq:genop}) and (\ref{eq:genen}) and introduce methods for efficient posterior
simulation in Section~\ref{sec:mcmc}.

\section{Sampling the posterior}\label{sec:mcmc}
While the posterior density function for $\beta$ given $\sigma^2$, $\lambda_1$, and $\lambda_2$
within a given orthant is multivariate normal, the 
orthant truncations create difficulties for direct analysis of the posterior as a whole due to
the potentially large number ($2^p$) of truncation components and the need to evaluate the
$p$-dimensional normal orthant probabilities $\mbox{P}(z, \mu^*_z, \Sigma^*_z)$ in the weights
$\omega^*_z$.  Fortunately, a simple MCMC algorithm can be constructed to obtain samples from
the posterior distribution $p(\beta \mid y, \sigma^2, \lambda_1, \lambda_2)$.  For the general
case under prior (\ref{eq:genen}), the full conditional distribution for $\beta_j$ is
\begin{equation}
	p(\beta_j \mid \beta_{-j}, y, \sigma^2, \lambda_1, \lambda_2) = (1-\phi_j) \mbox{N}^{-}
		(\beta_j \mid \mu_j^-, s^2_j) + \phi_j \mbox{N}^+(\beta_j \mid \mu_j^+, s^2_j),
		\label{eq:fullcon}
\end{equation}
where $\beta_{-j}$ is the vector of regression coefficients omitting element $j$, and $\mbox{N}^-$
and $\mbox{N}^+$ are density functions for truncated normal distributions as described
in Section~\ref{sec:iid}. The parameters $s^2_j$,
$\mu^+_j$, $\mu^-_j$ and $\phi_j$ of
the full conditional distributions arise from the usual calculations for normal distributions.  The conditional
scale parameters are given by
\[
	s^2_j = [ ({\Sigma^*}^{-1})_{jj}]^{-1} = \frac{\sigma^2}{\lambda_2 (\Sigma^{-1})_{jj} + x_j^T x_j},
\]
where the notation $(\cdot)_{jj}$ denotes the $j$th diagonal element of a matrix and $x_j$ is the 
$j$ column of the centered $X$ matrix.
The location parameters are
\begin{eqnarray}
	\mu^+_j &=& \mu^{*+}_j+ (\mu^{*+}_{-j} - \beta_{-j})^T ({\Sigma^*}^{-1})_{-j,j} / ({\Sigma^*}^{-1})_{jj}, 
		\label{eq:mupgen} \\
	\mu^-_j &=& \mu^{*-}_j+ (\mu^{*-}_{-j} - \beta_{-j})^T ({\Sigma^*}^{-1})_{-j,j} / ({\Sigma^*}^{-1})_{jj},
		\label{eq:mumgen}
\end{eqnarray}
where $({\Sigma^*}^{-1})_{-j,j}$ is the $j$th column of ${\Sigma^*}^{-1}$ with the $j$th row removed, and
$({\Sigma^*}^{-1})_{jj}$ is the $j$th diagonal entry of ${\Sigma^*}^{-1}$.  The ``$+$'' in $\mu^{*+}_j$ and
$\mu^{*+}_{-j}$ and the ``$-$'' in $\mu^{*-}_j$ and $\mu^{*-}_{-j}$ indicate the value of $z$ used to evaluate
$\mu^*$ in (\ref{eq:genpostmu}).  For example, $\mu^{*+}_j$ is the $j$th element of $\mu^*$, where $\mu^*$
is evaluated using $z_j = +1$ and $z_k = \mbox{sign}(\beta_k)\times 1$ for $k \neq j$.   Similarly,
$\mu^{*-}_j$ is the $j$th element of $\mu^*$, where $\mu^*$ is evaluated using $z_j = -1$ and 
$z_k = \mbox{sign}(\beta_k)\times 1$ for $k \neq j$.  The parameters $\mu^{*+}_{-j}$ and $\mu^{*-}_{-j}$ are 
calculated in the
same way and represent the resulting $(p-1)\times 1$ vectors where the $j$th element has been removed.
Finally, the weight $\phi_j$ for the non-negative component of (\ref{eq:fullcon}) is
\[
	\phi_j = \left\{ \frac{\Phi(\mu^+_j/s_j)}{\mbox{N}(0 \mid \mu^+_j, s^2_j)}\right\}
	\Big/\left\{ \frac{\Phi(\mu^+_j/s_j)}{\mbox{N}(0 \mid \mu^+_j, s^2_j)}
	+ \frac{\Phi(-\mu^-_j/s_j)}{\mbox{N}(0 \mid \mu^-_j, s^2_j)}
	\right\}.
\]

The expressions for $\mu^+_j$ and $\mu^-_j$ given in (\ref{eq:mupgen}--\ref{eq:mumgen}) are not 
computationally efficient, as they imply the need to invert the $\Sigma^*$ matrix at each step in the 
MCMC algorithm. We can re-express these parameters in terms of $\hat{\beta}_{G_{\lambda_2}}$ as
\[
	\mu_j^+ = \hat{\beta}_{{G_{\lambda_2}}_j} + 
		\left[ \sum_{i\neq j} \left(\hat{\beta}_{{G_{\lambda_2}}_i} - \beta_i \right)
		\frac{x_i^T   + \lambda_2 \Omega_{ij}}{x_j^T x_j + \lambda_2 \Omega_{jj}}
		\right]  - \frac{\lambda_1}{2(x_j^T x_j + \lambda_2 \Omega_{jj})},
\]
where $\Omega = \Sigma^{-1}$ and $(-\lambda_1)$ replaces $\lambda_1$ in the expression for
$\mu_j^-$. The parameter $\hat{\beta}_{G_{\lambda_2}}$ can be computed
by solving a system of linear equations. An even simpler expression is
\begin{equation}
	\mu_j^+ = \frac{x_j^T y - (x_j^T X_{-j} + \lambda_2 \Omega_{j,-j})\beta_{-j} - 
		\lambda_1/2}{x_j^T x_j + \lambda_2 \Omega_{jj}},
		\label{eq:mujcomp}
\end{equation}
where $X_{-j}$ is the centered $X$ matrix with the $j$th column removed, $\Omega_{j,-j}$ is the 
$j$th row of $\Omega$ with the $j$th column removed, and $(-\lambda_1)$ replaces $\lambda_1$ in the
expression for $\mu_j^-$. The terms $X^T X$, $X^T y$, and $\Omega$ can
be precomputed and are static at each step in the MCMC algorithm.

The specific case of the $\ell_1$-regularized $g$ prior (\ref{eq:l1g}) has $\Sigma = (X^TX)^{-1}$, 
which simplifies the expressions for $s^2_j$, $\mu^+_j$ and $\mu^-_j$:
\begin{eqnarray*}
	s^2_j &=& \frac{\sigma^2}{(1 + \lambda_2)x_j^T x_j}, \\
	\mu^+_j &=& \frac{1}{1+\lambda_2}\bolsj{j}
	+ \left[ \sum_{i \neq j}\left(\frac{1}{1+\lambda_2}\bolsj{i}
		- \beta_i\right)\frac{x_i^Tx_j}{x_j^Tx_j}\right] - \frac{\lambda_1}{2(1+\lambda_2)x_j^T x_j},
\end{eqnarray*}
where $\bolsj{j}$
is the $j$th element of the ordinary least squares estimate $\bols$ and $(-\lambda_1)$ replaces
$\lambda_1$ in the expression for $\mu_j^-$.

Simulating from the full conditional is straightforward. First, sample a Bernoulli random variable with
probability $\phi_j$. If the outcome is a success, sample $\beta_j$ from the truncated normal distribution
$\mbox{N}^+(\beta_j \mid \mu^+_j, s^2_j)$, otherwise sample $\beta_j$ from 
$\mbox{N}^-(\beta_j \mid \mu^-_j, s^2_j)$.  Efficient methods for sampling from univariate truncated
normal distributions are given in \citet{gewe:91}.

\subsection{Inference on $\sigma^2$, $\lambda_1$ and $\lambda_2$}\label{sec:fullbayes}
In treatments of Bayesian elastic net regression, $\sigma^2$ is often assigned an inverse gamma distribution, 
$\mbox{IG}(\nu_a/2, \nu_b/2)$, or its improper limit with $p(\sigma^2) \propto 1/\sigma^2$.
\citet{hans:11} assumed independent, rate-parameterized gamma priors $\lambda_1 \sim \mbox{Gamma}(L, \nu_1/2)$ 
and $\lambda_2 \sim \mbox{Gamma}(R, \nu_2/2)$ with mutual
independence among $\sigma^2$, $\lambda_1$, and $\lambda_2$. This prior is attractive when reformulating
the elastic net penalty function as $\lambda(\alpha \beta^T\beta + (1-\alpha)|\beta|_1)$ so that
$\lambda = \lambda_1 + \lambda_2$ is the overall penalization and $\alpha = \lambda_2/(\lambda_1 + \lambda_2)$
is the proportion of the total penalty attributable to the $\ell_2$-norm component. 
When $\nu_1 = \nu_2 = \nu$, the induced priors on the transformed parameters are
$\alpha \sim \mbox{Beta}(R,L)$ and $\lambda \sim \mbox{Gamma}(L+R, \nu/2)$,
independently of each other. The parameters $L$ and $R$ can be chosen to induce relatively more (or less)
penalization from either component, while the uniform prior with $L=R=1$ represents prior indifference.
Working with a different parameterization of the
Bayesian elastic net, \citet{wang:23} assumed prior distributions for a transformation of $\lambda_1$ and
$\lambda_2$ inspired by the use of the half-Cauchy distribution as a prior for scale \citep{gelm:06, pols:12b} 
and penalty parameters \citep{carv:10, bhad:17}. Any of these priors could be used for the generalized orthant
normal prior. We focus discussion in this paper on what we refer to as the ``base prior'':
\begin{equation}
	p_0(\sigma^2, \lambda_1, \lambda_2) = \mbox{IG}(\sigma^2 \mid \nu_a/2, \nu_b/2) \times
		\mbox{Gamma}(\lambda_1 \mid L, \nu_1/2) \times \mbox{Gamma}(\lambda_2 \mid R, \nu_2/2).
		\label{eq:base}
\end{equation}

In principle, posterior inference on $\sigma^2$, $\lambda_1$, and $\lambda_2$ is straightforward: one simply expands 
the model to include a prior distribution $p(\sigma^2, \lambda_1, \lambda_2)$ as above and adds appropriate sampling 
steps to the Gibbs sampler described above. 
In practice, posterior
sampling under the generalized orthant normal prior is complicated by the fact that the normalizing constant $\omega$ 
in the prior for $\beta$ (\ref{eq:genen}) is a non-trivial function of $\sigma^2$, $\lambda_1$ and $\lambda_2$:
\begin{eqnarray}
	\omega &=& \sum_{z \in \mathcal{Z}} \frac{\mbox{P}(z, \mu_z, \Sigma_z)}{\mbox{N}(0 \mid \mu_z, \Sigma_z)} 
		\nonumber \\
	&=&\sum_{z \in \mathcal{Z}} \frac{\mbox{P}\left(z, -\frac{\lambda_1}{2\lambda_2}\Sigma z, 
		\frac{\sigma^2}{\lambda_2}\Sigma\right)}{\mbox{N}\left(0 \mid -\frac{\lambda_1}{2\lambda_2}\Sigma z,
		\frac{\sigma^2}{\lambda_2}\Sigma\right)} \label{eq:sumz} \\
	&\equiv& \omega(\sigma^2, \lambda_1, \lambda_2), \nonumber
\end{eqnarray}
where the notation in the final expression emphasizes the dependence of the normalizing constant $\omega$ 
on the parameters.

Under the base prior (\ref{eq:base}) and other commonly-used priors, the full conditional density functions for 
$\sigma^2$, $\lambda_1$, and $\lambda_2$ will all involve the term $\omega(\sigma^2, \lambda_1, \lambda_2)$
and will not be ``standard'' distributions from which obtaining samples is easy. Updating any of these parameters in 
an MCMC scheme via a Metropolis--Hastings update would require evaluation of $\omega(\sigma^2, \lambda_1, \lambda_2)$ 
under both the current and proposed value(s) of the parameter(s) being updated.  Evaluating expression (\ref{eq:sumz}) 
directly as written requires the evaluation of $2^p$ integrals of multivariate normal densities, each restricted to a 
different orthant in $\mathbb{R}^p$.  The two obvious computational difficulties in evaluating 
$\omega(\sigma^2, \lambda_1, \lambda_2)$ as written in (\ref{eq:sumz}) are the number of terms in the sum 
and computation of the normal probabilities.  Numerical routines for estimating multivariate normal 
probabilities such as the \texttt{pmvnorm} function in the \texttt{R} package \texttt{mvtnorm} of \citet{genz:12} 
\citep[see also][]{genz:92, genz:93, genz:09}  can be used to evaluate
$\omega(\sigma^2, \lambda_1, \lambda_2)$ when $p$ is not too large; however, this direct approach breaks down
in practice when $p$ is large due to the number of components in the sum, the high dimensionality of the 
integrals, and the fact that the evaluations must be done whenever any of the three parameters are updated in 
the MCMC sampler.

We can simplify the expression for $\omega(\sigma^2, \lambda_1, \lambda_2)$ by writing the prior density for 
$\beta$ as
\begin{eqnarray*}
	p(\beta \mid \sigma^2, \lambda_1, \lambda_2) &=& \sum_{z \in \mathcal{Z}} \omega_z \mbox{N}^{[z]}
		\left( \beta \left| -\frac{\lambda_1}{2\lambda_2}\Sigma z, \frac{\sigma^2}{\lambda_2}\Sigma\right. \right) \\
	&=& \omega(\sigma^2, \lambda_1, \lambda_2)^{-1} \exp\left\{ -\frac{1}{2\sigma^2}\left( \lambda_2 \beta^T 
		\Sigma^{-1} \beta + \lambda_1 |\beta|_1\right)\right\}
\end{eqnarray*}
(see Appendix~\ref{app:densprop}).
The normalizing constant therefore satisfies
\begin{eqnarray}
	\omega(\sigma^2, \lambda_1, \lambda_2) &=& \int \exp\left\{ -\frac{1}{2\sigma^2}\left( \lambda_2 \beta^T 
		\Sigma^{-1} \beta + \lambda_1 |\beta|_1\right)\right\} d\beta \nonumber \\
		&=& (2\pi)^{p/2} \left(\frac{\sigma^2}{\lambda_2}\right)^{p/2} |\Sigma|^{1/2}\; \mathbb{E}\left[
			e^{-\frac{\lambda_1}{2\sigma \sqrt{\lambda_2}}|Z|_1} \right], \label{eq:omegaEV}
\end{eqnarray}
where the random variable in the expectation is $Z \sim \mbox{N}(0, \Sigma)$. Rather than computing
$2^p$ individual $p$-dimensional multivariate normal probability calculations, we need instead compute 
the expected value of a function of the $\ell_1$-norm of a multivariate normal random variable. When
$\Sigma$ is a diagonal matrix with diagonal elements $v^2_j$, $j=1, \ldots, p$, the expectation in 
(\ref{eq:omegaEV}) can be written in terms of the standard normal cdf, $\Phi(\cdot)$:
\[
	\mathbb{E}\left[e^{-\frac{\lambda_1}{2\sigma \sqrt{\lambda_2}}|Z|_1} \right] =
		2^p \exp\left\{\frac{\lambda_1^2}{8\sigma^2\lambda_2}\sum_{j=1}^p v_j^2\right\}
		\prod_{j=1}^p \Phi\left(-\frac{\lambda_1}{2\sigma\sqrt{\lambda_2}}v_j\right),
\]
which can be computed to high accuracy unless the argument to $\Phi$ is very large.
When $v^2_j = 1$ for all $j$, the expression matches the term in the normalizing constant for
Bayesian elastic net regression \citep[see][]{hans:11}.

For general $\Sigma$, there are no simple expressions for the expected value. The form of
(\ref{eq:omegaEV}) suggests that we might consider Monte Carlo approaches to estimating
$\omega(\sigma^2, \lambda_1, \lambda_2)$. A na\"ive approach might sample $Z$ directly
from a $\mbox{N}(0, \Sigma)$ distribution and estimate the expectation via an empirical
average; more sophisticated approaches might use importance sampling to reduce the
variance of the resulting estimate. 
If the resulting estimates of $\omega(\sigma^2, \lambda_1, \lambda_2)$
are extremely accurate, we might be comfortable simply plugging them in when computing
acceptance probabilities in a Metropolis--Hastings update for $\sigma^2$, $\lambda_1$,
or $\lambda_2$. In reality, even a small amount of error might cause the
resulting chain to either (i) appear to converge to the correct distribution when it
hasn't, or (ii) fail to converge entirely, with sample paths diverging to $0$ or $\infty$.
We have observed the latter behavior for some data sets when using the na\"ive Monte
Carlo approach, even for very large Monte Carlo samples. With no clear method for 
estimating $\omega(\sigma^2, \lambda_1, \lambda_2)$
accurately in hand, we pursue another approach for working with the model.

\subsection{Computationally tractable priors and posteriors}\label{sec:newprior}
To avoid the computational issues associated with the term $\omega(\sigma^2, \lambda_1, \lambda_2)$
in the normalizing constant for the prior on $\beta$, we propose using the following prior
distribution for $\sigma^2$, $\lambda_1$, and $\lambda_2$ for full Bayesian inference:
\begin{eqnarray}
	p(\sigma^2, \lambda_1, \lambda_2) &\propto& \mathbb{E}\left[e^{-\frac{\lambda_1} 
		{2\sigma\sqrt{\lambda_2}}|Z|_1}\right] \times p_0(\sigma^2, \lambda_1, \lambda_2) \nonumber \\
	&=& \mathbb{E}\left[e^{-\frac{\lambda_1}{2\sigma\sqrt{\lambda_2}}|Z|_1}\right] \times \nonumber \\
	&& \mbox{IG}\left(\sigma^2 \mid \frac{\nu_a}{2}, \frac{\nu_b}{2}\right) \times
			\mbox{Gamma}\left(\lambda_1 \mid L, \frac{\nu_1}{2}\right) \times 
	\mbox{Gamma}\left(\lambda_2 \mid R, \frac{\nu_2}{2}\right),
			\label{eq:newprior}
\end{eqnarray}
where $p_0(\cdot)$ is the base prior density (\ref{eq:base}) and $Z \sim \mbox{N}(0,\Sigma)$. 
The new prior is proper because
$0 < \mathbb{E}\left[e^{-\frac{\lambda_1}{2\sigma\sqrt{\lambda_2}}|Z|_1}\right] < 1$ 
for all $\sigma^2  > 0$, $\lambda_1 > 0$, and $\lambda_2 > 0$ and the base prior is proper.
Importantly, the intractable integral now cancels out in the joint prior on all model parameters:
\begin{eqnarray}
	p(\beta, \sigma^2, \lambda_1, \lambda_2) &\propto&
	(\sigma^2)^{-(p+\nu_a)/2 - 1} \lambda_1^{L-1} \lambda_2^{p/2 + R - 1} \times 
	\nonumber \\
	&& \exp\left\{-\frac{1}{2\sigma^2}\left(\lambda_2 \beta^T \Sigma^{-1}\beta + \lambda_1 |\beta|_1
		+ \nu_b\right) - \lambda_1\nu_1/2 - \lambda_2\nu_2/2 \right\}.  \label{eq:newjoint}
\end{eqnarray}
We have effectively modulated the base prior used by \citet{hans:11} via the term 
$\mathbb{E}\left[e^{-\frac{\lambda_1}{2\sigma\sqrt{\lambda_2}}|Z|_1}\right]$ to produce
a new prior that combines with $p(\beta \mid \sigma^2, \lambda_1, \lambda_2)$ to yield a
joint prior (\ref{eq:newjoint}) that is computationally tractable and results in a 
computationally tractable posterior (see below).

\subsubsection{\emph{Caveat} sampler} \label{sec:caveat}
When arbitrarily modifying a probability model to facilitate posterior sampling, one
should beware of the possibility of unintended impacts on inference. For example, we
might be concerned that we have modified the tail behavior of the prior in an 
unexpected way. We might also be concerned that, if the original prior reasonably
reflected out prior beliefs marginally for each parameter, modulating the joint prior
might drastically alter the original margins. We investigate these concerns by
(i) marginalizing $\lambda_1$ and $\lambda_2$ from the joint
prior and inspecting the resulting prior for $\beta$ given $\sigma^2$,
and (ii) simulating from the joint prior distribution (\ref{eq:newjoint}) in order to 
visualize the marginal priors on each parameter. The former is useful for understanding 
the analytical structure of the new prior, while the latter is useful in practice when 
assigning a prior for a particular data analysis.

Integrating (\ref{eq:newjoint}) with respect to $\lambda_1$ and/or $\lambda_2$ yields
\begin{eqnarray*}
	p(\beta \mid \sigma^2, \lambda_2) &\propto& \exp\left\{-\frac{\lambda_2}{2\sigma^2}
		\beta^T \Sigma^{-1}\beta\right\} \times \left(1 + \frac{|\beta|_1}{\nu_1 \sigma^2}
		\right)^{-L}, \\
		p(\beta \mid \sigma^2, \lambda_1) &\propto& \left(1 + \frac{\beta^T \Sigma^{-1}\beta}
		{\nu_2 \sigma^2}\right)^{-(2R + p)/2} \times
	\exp\left\{-\frac{\lambda_1}{2\sigma^2}
		|\beta|_1\right\},  \\
	p(\beta \mid \sigma^2) &\propto& \left(1 + \frac{\beta^T \Sigma^{-1}\beta}
		{\nu_2 \sigma^2}\right)^{-(2R + p)/2} \times \left(1 + \frac{|\beta|_1}
		{\nu_1 \sigma^2}\right)^{-L}.
\end{eqnarray*}
In each case we obtain a novel prior with either exponential tails (when marginalizing over
only one of the penalty parameters) or polynomial tails (when marginalizing over both).
An interesting case arises when $R = \nu_2/2$, which results in
\[
	p(\beta \mid \sigma^2) \propto t_{\nu_2}(\beta \mid 0, \sigma^2 \Sigma) \times
		\left(1 + \frac{|\beta|_1}{\nu_1 \sigma^2}\right)^{-L}.
\]
The prior density is a product of the density function for a multivariate $t$-distribution 
and a function that is similar to the density function for a multivariate $t$-distribution 
with an $\ell_1$-norm replacing the squared $\ell_2$-norm. When $p=1$, a special case of the 
latter is known as a double Lomax distribution \citep{puna:11, puna:15}. \citet{gu:13} explored
the use of independent double Lomax prior distributions in sparse regression problems.
When $L = (\nu_2 + p)/2$, the exponents for the two components of the density function
match, and we can write
\begin{equation}
	-2 \log p(\beta \mid \sigma^2) = \mbox{const. } + (\nu_2 + p)\left[
		\log \left(1 + \frac{\beta^T\Sigma^{-1}\beta}{\nu_2 \sigma^2}\right)
		+ \log \left(1 + \frac{|\beta|_1}{\nu_1 \sigma^2}\right)
		\right]. \label{eq:l1t}
\end{equation}
Matching the exponents allows the relative strengths of the two components of the
polynomial tails to be controlled by size of $\nu_1$ relative to $\nu_2$, with smaller
values of $\nu_1$ corresponding to the $\ell_1$-norm component contributing more heavily.
This form of the prior is an attractive choice because it reduces the number of hyperparameters
that need to be specified. It also allows $L$ to scale with $p$, automatically inducing
a stronger peak in the prior density at the origin in higher-dimensional problems. We
refer to this form (\ref{eq:l1t}) of the prior as the ``$\ell_1$-regularized $t$ prior.''

Modulating the base prior by the term 
$\mathbb{E}\left[e^{-\frac{\lambda_1}{2\sigma\sqrt{\lambda_2}}|Z|_1}\right]$ induces
prior dependence among $\sigma^2$, $\lambda_1$, and $\lambda_2$ and results in
marginal priors that are not necessarily inverse gamma or gamma distributions. To
facilitate prior specification for a given data analysis, we suggest using a Gibbs
sampler to sample from the joint prior distribution (\ref{eq:newjoint}) and 
plotting estimates of the marginal (or joint) prior densities of $\sigma^2$,
$\lambda_1$, and $\lambda_2$ to assess whether 
specific values of $\nu_a$, $\nu_b$, $L$, etc., result in a prior that is consistent 
with prior beliefs.  Samples from the joint prior can be obtained as follows.
\begin{enumerate}
\item Iteratively for $j=1, \ldots, p$, sample 
$\beta_j \mid \beta_{-j}, \sigma^2, \lambda_1, \lambda_2$ with probability
$\phi_{0j}$ from a positively truncated normal distribution with location $\mu_{0j}^+$
and scale $s^2_{0j}$ and with probability $1-\phi_{0j}$ from a negatively truncated
normal distribution with location $\mu_{0j}^-$ and scale $s^2_{0j}$, where
$s_{0j}^2 = \sigma^2/(\lambda_2 \Omega_{jj})$,
$
	\mu_{0j}^{\pm} = (-\Omega^T_{-j,j}\beta_{-j} \mp 
	\lambda_1 /(2\lambda_2))\Omega_{jj}^{-1},
$
\[
	\phi_{0j} = \frac{\Phi(\mu_{0j}^+/s_{0j})}{\mbox{N}(0 \mid \mu_{0j}^+,
		s^2_{0j})} \Big/ \left\{ \frac{\Phi(\mu_{0j}^+/s_{0j})}{\mbox{N}(0 \mid \mu_{0j}^+,
		s_{0j}^2)} + \frac{\Phi(-\mu_{0j}^-/s_{0j})}{\mbox{N}(0 \mid \mu_{0j}^-,
		s_{0j}^2)}\right\},
\]
$\Omega = \Sigma^{-1}$, and the expressions $\Omega_{jj}$ and $\Omega_{-j,j}$ are
as defined in Section~\ref{sec:mcmc}.
\item Sample $\sigma^2 \mid \beta, \lambda_1, \lambda_2 \sim 
\mbox{IG}\left(\frac{\nu_a + p}{2}, 
\frac{\lambda_1 |\beta|_1 + \lambda_2 \beta^T \Sigma^{-1} \beta + \nu_b}{2} \right)$.
\item Sample $\lambda_1 \mid \beta, \sigma^2, \lambda_2 \sim \mbox{Gamma}\left(
L, \frac{|\beta|_1/\sigma^2 + \nu_1}{2}\right)$. \label{it:lam1fullcon}
\item Sample $\lambda_2 \mid \beta, \sigma^2, \lambda_1 \sim \mbox{Gamma}\left(
R + p/2, \frac{\beta^T \Sigma^{-1} \beta/\sigma^2 + \nu_2}{2}\right)$.
\label{it:lam2fullcon}
\end{enumerate}
The density functions for all full conditional prior distributions can be computed
numerically and so Rao--Blackwellized estimates of the marginal densities can be
easily computed using the sampled parameter values. We illustrate this approach
for visualizing the prior in an example in Section~\ref{sec:illustration}.

\subsubsection{Sampling from the posterior}\label{sec:newpost}
Sampling from the full posterior distribution via Gibbs sampling is straightforward 
under prior (\ref{eq:newprior}). The regression coefficients $\beta_j$ are sampled from
their full conditional distributions as described in Section~\ref{sec:mcmc}.
The penalty parameters $\lambda_1$ and $\lambda_2$ are sampled from their full
conditional distributions, which, due to the structure of the model, 
are the same in the posterior as in the prior. These distributions are given in 
steps~\ref{it:lam1fullcon} and \ref{it:lam2fullcon} for sampling from 
the prior in Section~\ref{sec:caveat}. The variance of the error term is sampled from 
its full conditional distribution,
\[
	\sigma^2 \mid y, \beta, \lambda_1, \lambda_2 \sim
	\mbox{IG}\left(\frac{n+p+\nu_a-1}{2},
		\frac{(y - X\beta)^T(y - X\beta) + \lambda_2 \beta^T \Sigma^{-1} \beta +
			\lambda_1 |\beta|_1 + \nu_b}{2}\right).
\]

\section{Illustration}\label{sec:illustration}
We illustrate the $\ell_1$-regularized $g$ prior using the prostate cancer data of 
\citet{stam:89}, obtained from the \texttt{R} package \texttt{lasso2} \citep{lokh:07}. 
The dependent variable is the logarithm of prostate-specific antigen for each of $n=97$ 
patients. There are $p=8$ clinical variables available as predictors of the dependent variable.
This data set has been used to illustrate other regularized regression methods in previous
work \citep[e.g.,][]{tibs:96, fu:98}. In this section we work with the eight predictors as 
specified and used in those previous examples (\texttt{lcavol}, \texttt{lweight}, 
\texttt{age}, \texttt{lbph}, \texttt{svi}, \texttt{lcp}, \texttt{gleason}, \texttt{pgg45}) 
for consistency. 
For the illustration, $y$ and all predictors $x_j$ 
are mean-centered as described in Section~\ref{sec:onpost}.  The predictor variables are 
additionally transformed so that $ x_j^T x_j = n-1$ (i.e., $s^2_{x_j} = 1$) so that the 
coefficients share a common scale. 

\subsection{Illustration 1: Visualizing the prior}\label{sec:visprior}
It is tempting to select hyperparameter values that reflect prior beliefs
about $\sigma^2$, $\lambda_1$, and $\lambda_2$ under the base prior (\ref{eq:base}),
modulate the prior to obtain the computationally tractable prior (\ref{eq:newprior}),
and then proceed directly with posterior inference. It is important, however, to 
check to ensure that the modulation hasn't impacted the prior too strongly or 
in a way that is inconsistent with the information we wish to include in the analysis.
We illustrate the effect of the modulation by using the Gibbs sampling approach described
in Section~\ref{sec:caveat} to visualize the base and modulated priors under the following
settings for the hyperparameters $L$, $R$, $\nu_1$, and $\nu_2$:
\begin{description}
\item[Base-uniform:] $L = R = 1$, $\nu_1 = \nu_2 = 2$. This corresponds to a uniform
prior on $\alpha = \lambda_2/(\lambda_1 + \lambda_2)$ under the base prior.
\item[L1-t:] The $\ell_1$-regularized $t$ prior with $L = (\nu_2 + p)/2$,
$R = \nu_2/2$, and $\nu_1 = \nu_2 = 2$.
\end{description}
In both cases, we set $\nu_a = 10$ and $\nu_b = 4$. Under the base prior, this would
correspond to an inverse gamma prior on $\sigma^2$ with mean $0.5$. The classical
unbiased estimate of $\sigma^2$ under this scaling of the data is 
$\hat{\sigma}^2 = 0.496$, and so a base prior with these values of $\nu_a$ and
$\nu_b$ might reflect the prior beliefs of a well-informed subject-matter 
expert.

\begin{figure}[!ht]
 \begin{center}
 \includegraphics[scale=0.5]{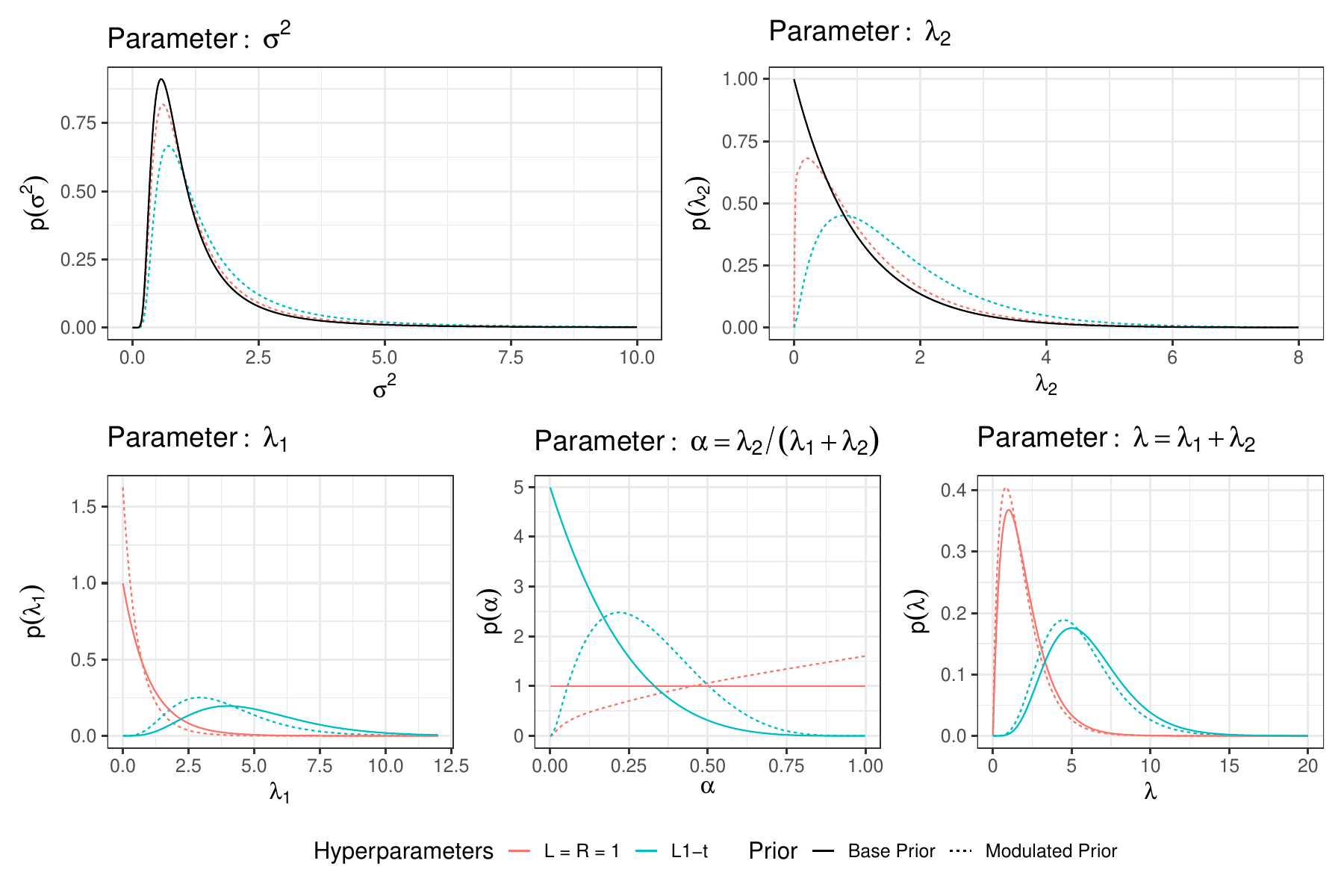}
 \caption{Rao--Blackwellized estimates of the marginal densities under the modulated
 prior (dashed lines) for two different hyperparameter settings. The corresponding
 base priors are shown as solid lines for comparison. Black lines correspond to
 cases where the base prior is the same under both hyperparameter settings. For
 all priors, $\nu_a = 10$, $\nu_b = 4$, and $\nu_1 = \nu_2 = 2$.
 	\label{fig:base-mod}}
 \end{center}
\end{figure}

Setting $\Sigma = (X^TX)^{-1}$, we obtained 10,000 samples from the joint prior 
distribution (\ref{eq:newjoint}) using the Gibbs sampler described in 
Section~\ref{sec:caveat} under both hyperparameter settings. Figure~\ref{fig:base-mod}
displays Rao--Blackwellized density estimates of the marginal prior densities for
$\sigma^2$, $\lambda_1$, and $\lambda_2$ (dashed lines). The density estimates are
constructing using the MCMC samples and the full conditional prior densities
described in Section~\ref{sec:caveat}. 
In addition to these three parameters, we also visualize the prior distributions for 
$\alpha = \lambda_2/(\lambda_1 + \lambda_2)$ and $\lambda = \lambda_1 + \lambda_2$. 
Rao--Blackwellized estimates of the marginal prior densities
under the base and modulated priors are shown in Figure~\ref{fig:base-mod}.
The density estimates for $\alpha$ and $\lambda$
are constructed by transforming the MCMC samples of $\lambda_1$ and $\lambda_2$
and using the following full conditional densities:
\begin{eqnarray*}
	\lambda \mid \beta, \sigma^2, \alpha &\sim& \mbox{Gamma}\left(
		p/2 + R + L, \frac{\alpha\left(\nu_2 + \beta^T \Sigma^{-1}\beta/\sigma^2\right)
			+ (1-\alpha)\left(\nu_1 + |\beta|_1\sigma^2\right)}{2}\right) \\
	p(\alpha \mid \beta, \sigma^2, \lambda) &=& \frac{\alpha^{p/2+R-1}(1-\alpha)^{L-1}
		\exp\left\{-\alpha \, s(\lambda, \beta, \sigma^2)\right\}}
		{\mbox{B}(p/2+R,L) _1\mbox{F}_1\left(p/2+R, p/2+R+L, -s(\lambda, \beta, \sigma^2)\right)},
\end{eqnarray*}
where $s(\sigma^2, \lambda_1, \lambda_2) = 
	\lambda\left[(\beta^T\Sigma^{-1}\beta - |\beta|_1)/\sigma^2 + \nu_2 - \nu_1\right]$.
The full conditional for $\alpha$ is a confluent hypergeometric distribution \citep{gord:98}. 
The beta function, $\mbox{B}$, and the $_1\mbox{F}_1$ function can be computed numerically in 
\texttt{R}, the latter using the \texttt{gsl} package \citep{hank:06}.
Finally, the marginal densities under the base prior are displayed in Figure~\ref{fig:base-mod} 
using solid lines. 

Focusing first on $\sigma^2$, the two modulated priors are slightly different from the
base prior due to the dependence between parameters in the modulated prior (the
base prior is the same under both hyperparameter settings), and the two modulated
priors are slightly different from each other. The mode of the base prior, though, is 
similar to the mode of both modulated priors, and so we can use the base prior as rough 
proxy for the modulated prior when specifying the hyperparameters $\nu_a$ and $\nu_b$. 
Though not clear from the plot, we note that both modulated priors for $\sigma^2$ have 
heavier tails than the base inverse gamma prior.

We observe a larger difference between the base and modulated priors for $\lambda_2$.
The base prior for $\lambda_2$ is an exponential distribution with mean 1, while
the marginals under the modulated prior have modes that are shifted toward larger
values. Less impacted by the modulation are the marginal priors on $\lambda_1$,
which are similar to the corresponding base priors under both hyperparameter settings.
Interestingly, the same is true for $\lambda = \lambda_1 + \lambda_2$, despite
the differences between the base and modulated marginal priors for $\lambda_2$.

The modulation of the base prior has a substantial impact on the distribution of
$\alpha$, the fraction of the total penalty that is attributed to the $\ell_2$-norm
component of the prior. The base prior on $\alpha$ is uniform when $L=R=1$ and
$\nu_1=\nu_2$, however we see that the modulated prior under this hyperparameter
setting is not uniform, with nearly linearly increasing density function. This
suggests that, if we wish to maintain a uniform prior on $\alpha$, we need to 
be mindful of the impact of the modulation on the joint prior distribution
and experiment with different hyperparameter settings. In this example, setting 
$R = 0.5$ results in a modulated prior on $\alpha$ that is nearly uniform over most 
of its domain. The effect of the modulation is also strong under the L1-t prior. 
Smaller values of $\alpha$ correspond to stronger relative $\ell_1$-norm
penalization. Under the L1-t prior, the modulated prior on $\alpha$ favors
small values of $\alpha$ slightly less than does the base prior, but still more
so than when $L=R=1$.

In this example, we see that modulation of the base prior by the term
$\mbox{E}\left[e^{-\frac{\lambda_1}{2\sigma\sqrt{\lambda_2}}|Z|_1}\right]$ results
in a prior that facilitates computation while maintaining the general shape of the
marginal priors for $\sigma^2$ and $\lambda_1$ (and, to a lesser extent, $\lambda_2$).
Simulating from and visualizing the modulated prior is helpful step for ensuring
our prior beliefs are accurately incorporated into the model for a specific data analysis.

\renewcommand{\baselinestretch}{1}
\renewcommand{\baselinestretch}{1.85}

\subsection{Illustration 2: Comparison with the $g$ prior}\label{sec:fixedg}
We next illustrate the $\ell_1$-regularized $g$ prior by comparing it to Zellner's $g$ prior
with $g = n$, which corresponds to a unit information prior \citep{kass:95, fern:01}.
This comparison allows us to illustrate the impact of including the $\lambda_1 |\beta|_1$ term 
in the prior on the posterior distribution of the regression coefficients. 
The prior model under Zellner's $g$ prior is taken to be $\beta \mid \sigma^2, g \sim 
\mbox{N}(0, \sigma^2 n (X^TX)^{-1})$ and $\sigma^2 \sim \mbox{IG}(\nu_a/2, \nu_b/2)$. To 
make the comparison as direct as possible, we fix $\lambda_2 = n^{-1}$ in the $\ell_1$-regularized 
$g$ prior.  We then compute the posterior for three fixed values of $\lambda_1 \in \{1, 5, 15\}$, 
representing weak, moderate, and large amounts of $\ell_1$-norm penalization. The prior on $\sigma^2$
is constructed given $\lambda_1$ and $\lambda_2$ and is a modulated version of the prior
for $\sigma^2$ under Zellner's $g$ prior:
\[
	p(\sigma^2 \mid \lambda_1, \lambda_2 = n^{-1}) \propto
		\mathbb{E}\left[e^{-\frac{\lambda_1\sqrt{n}}{2\sigma}|Z|_1}\right] \times
		\mbox{IG}\left(\sigma^2 \mid \frac{\nu_a}{2}, \frac{\nu_b}{2}\right),
\]
where $Z \sim \mbox{N}(0, (X^TX)^{-1})$. 
We set $\nu_a = 10$ and $\nu_b = 4$ for both Zellner's $g$ prior and the $\ell_1$-regularized
$g$ prior. The model under Zellner's $g$ prior is identical to the $\ell_1$-regularized $g$ prior
with $\lambda_1=0$, and so small values of $\lambda_1$ should produce posteriors that are 
similar to Zellner's $g$ prior.

\begin{figure}[!t]
 \begin{center}
 \includegraphics[scale=0.5]{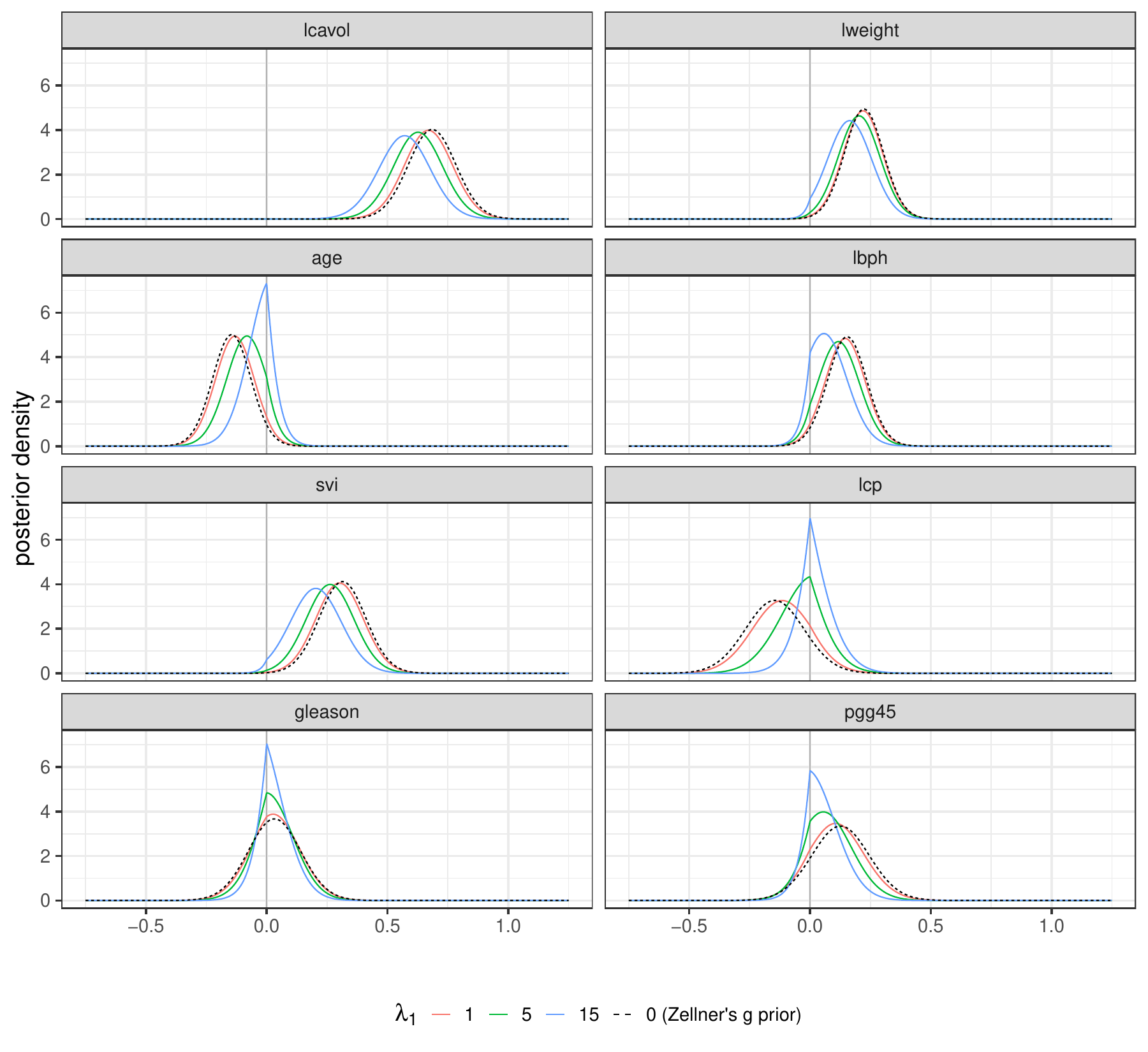}
 \caption{Illustration~2: Marginal posterior densities of the regression coefficients, $\beta_j$,
		labeled by predictor name under the
 		$\ell_1$-regularized $g$ prior for different fixed values of $\lambda_1$
		with $\lambda_2^{-1} = g = n$. Zellner's $g$ prior corresponds to the
		cases where $\lambda_1=0$.
 	\label{fig:l1g-zg}}
 \end{center}
\end{figure}

Sampling $\beta$ and $\sigma^2$ from the posterior under the $\ell_1$-regularized
$g$ prior proceeds as described in Section~\ref{sec:newpost} with $\lambda_2$ fixed at $n^{-1}$
and $\lambda_1$ fixed at one of the three specified values (ignoring the steps for sampling
these fixed parameters). We obtained 10,000 samples from the posterior after a burn-in of $100$ 
iterations. Figure~\ref{fig:l1g-zg} displays Rao--Blackwellized estimates of the marginal posterior 
densities of the $\beta_j$ under the $\ell_1$-regularized $g$ prior, 
$p(\beta_j \mid y, \lambda_1, \lambda_2 = n^{-1})$, for the three different values of $\lambda_1$.
The dashed black lines are the marginal posterior densities under Zellner's $g$ prior,
$p(\beta_j \mid y, g = n)$, which are all margins of the multivariate $t$ distribution
\[
	\beta \mid y, g \sim t_{n-1}\left( \frac{g}{1+g}\bols,
		\frac{g}{1+g} s^2_g (X^TX)^{-1} \right),
\]
where $s^2_g = (y^T(I_n - \frac{g}{1+g}P_X)y + \nu_b)/(n-1+\nu_a)$, $P_X = X(X^TX)^{-1}X^T$,
and $g=n$.

We see in Figure~\ref{fig:l1g-zg} that the marginal posterior distributions of the regression 
coefficients when $\lambda_1$ is small are nearly identical to those under Zellner's $g$ prior.
Though the posterior densities are not differentiable at zero when $\lambda_1 > 0$, we cannot
observe this phenomenon by eye when $\lambda_1 = 1$. We note a few specific behaviors as
$\lambda_1$ increases. The posteriors for three of the large regression coefficients 
under Zellner's $g$ prior (\texttt{lcavol}, \texttt{lweight}, and \texttt{svi}) are shrunk
toward zero, but even at large levels of $\ell_1$-norm penalization they still have a roughly
$t$-distribution-like shape with minimal visual non-differentiability at zero.
For the other coefficients, as $\lambda_1$ increases we observe both an increasing skew
in the marginal densities and a more pronounced non-differentiable peak at zero. This 
effect is most pronounced for the predictors \texttt{age}, \texttt{lcp}, \texttt{gleason}, 
and \texttt{pgg45}, which had modes close to zero under Zellner's $g$ prior. The two
predictors \texttt{lweight} and \texttt{lbph} provide an interesting contrast. While the
coefficients for both predictors have similar marginal posteriors under Zellner's $g$ prior,
the impact of the $\ell_1$-norm penalization is quite different as $\lambda_1$ increases.
The distribution of the coefficient for \texttt{lbph} is shrunk toward zero more rapidly
and has noticeable skew and non-differentiability when $\lambda_1=15$. The difference 
in the rate of shrinkage toward zero is due to the dependence among all predictors, which
shows up in the model in the $\frac{\lambda_1}{2}(X^TX)^{-1} z$ term in (\ref{eq:mupost}).

\renewcommand{\baselinestretch}{1}
\renewcommand{\baselinestretch}{1.85}

\subsection{Illustration 3: Full Bayesian Inference}\label{sec:full}
The third part of the illustration extends Illustration~2 in Section~\ref{sec:fixedg} by treating both 
$\lambda_1$ and $\lambda_2$ as unknown parameters using prior (\ref{eq:newjoint}). We compare the 
posterior under the fully-Bayes, $\ell_1$-regularized $g$ prior to the posterior under the Zellner--Siow 
prior \citep{zell:80} and the hyper-$g/n$ prior \citep{lian:08}. The Zellner--Siow prior is a mixture of 
ordinary $g$ priors with $g \sim \mbox{IG}(1/2, n/2)$, while the hyper-$g/n$ prior is a mixture of 
ordinary $g$ priors with $p(g) = \frac{a-2}{2n}(1 + g/n)^{-a/2}$ with $a > 2$ 
\citep[as recommended by][we use the value $a=3$]{lian:08}. The Zellner--Siow prior corresponds to a 
multivariate Cauchy distribution on $\beta$ given $\sigma^2$. To make the comparison as similar as 
possible, we set $R = 1/2$, $\nu_2 = n$, $L=(1+p)/2$, and $\nu_1 = n$ in the base prior. The marginal 
prior for $\beta$ given $\sigma^2$ is
\[
	p(\beta \mid \sigma^2) \propto t_1(\beta \mid 0, \sigma^2 n (X^TX)^{-1}) \times
		\left(1 + \frac{|\beta|_1}{n \sigma^2}\right)^{-(1+p)/2},
\]
an $\ell_1$-penalized Zellner--Siow prior. For all three models we set $\nu_a = 10$ and $\nu_b = 4$.

\begin{figure}[!t]
 \begin{center}
 \includegraphics[scale=0.5]{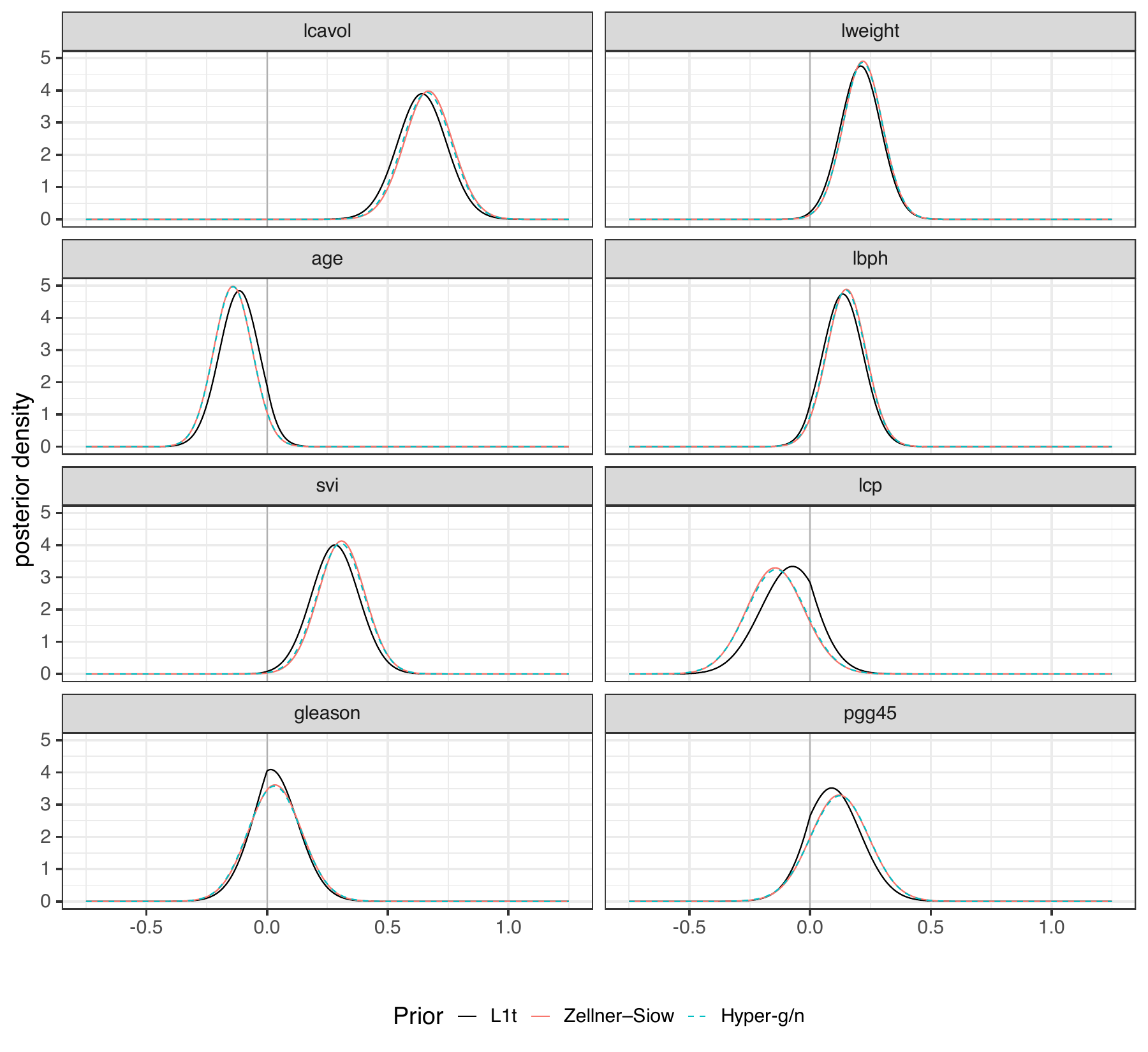}
 \caption{Illustration~3: Marginal posterior densities of the regression coefficients, $\beta_j$,
 labeled by predictor name under the $\ell_1$-regularized $g$ prior when both $\lambda_1$ and 
 $\lambda_2$ are assigned prior distributions. The posteriors under the Zellner--Siow and hyper-$g/n$ 
 priors are displayed for comparison (see Section~\ref{sec:full} for details).
 	\label{fig:density_full}}
 \end{center}
\end{figure}

Figure~\ref{fig:density_full} displays Rao--Blackwellized estimates of the marginal posterior densities of 
the $\beta_j$ under each of the three models. The Zellner--Siow and hyper-$g/n$ priors result in nearly
identical marginal posteriors for all regression coefficients. For the $\ell_1$-regularized $g$ prior, 
putting a prior on $\lambda_1$ smooths out the marginal posterior densities compared to the fixed-$\lambda_1$
case in Figure~\ref{fig:l1g-zg}, but the $\ell_1$-norm penalization is still apparent for several of
the smaller coefficients. The marginal posterior densities for the coefficients for \texttt{lcp}, 
\texttt{gleason}, and \texttt{pgg45} are all noticeably asymmetric and non-differentiable at zero. 
The posteriors for the other regression coefficients are more similar to the Zellner--Siow and 
hyper-$g/n$ posteriors, but are shrunk slightly more toward zero due to the additional penalty term.

\renewcommand{\baselinestretch}{1}
\renewcommand{\baselinestretch}{1.85}

\section{Simulation and predictive comparison}\label{sec:sim}
The relative performance of the Bayesian elastic net (with $\Sigma = I_p$) \emph{viz-\`a-viz}
other regularized regression methods has been well-documented in the literature 
\citep[e.g.,][]{li:10, hans:11,roy:17, wang:23}. Here, we focus on quantifying the potential
for improvement in predictive performance that can be achieved through judicious choice of
the dependence matrix, $\Sigma$, in the generalized orthant normal prior. We consider the 
four simulation setups used by \citet{zou:05} in their original study of the elastic net.
These settings, which have also been used by \citet{hans:11} and \citet{wang:23}, allow us
to incorporate prior dependence between the regression coefficients in the model based on
knowledge about patterns among the true regression coefficients. In each of the four settings,
we first simulate $X \sim \mbox{N}(0, V)$ and then simulate $y \sim \mbox{N}(X\beta, \sigma^2 I_n)$, 
where $n$, the $p\times p$ matrix $V$, the $p$-vector $\beta$, and $\sigma^2$ are specific to each 
simulation setting as described below.
\begin{description}
\item[Simulation Setting 1:] $n=20$, $p=8$, $\beta = (3, 1.5, 0, 0, 2, 0, 0, 0)^T$, $\sigma^2 = 9$,
$V_{ij} = (0.5)^{|i-j|}$, and $1 \leq i,j \leq p$.
\item[Simulation Setting 2:] Same as Setting~1, but with $\beta_j = 0.85$, $j=1,\ldots,p$.
\item[Simulation Setting 3:] $n=100$, $p=40$, $\sigma^2 = 225$, $V_{ii} = 1$, $i=1,\ldots,p$,
$V_{ij} = 0.5$ for all $i\neq j$, $\beta = (0, \ldots, 0, 2, \ldots, 2, 0, \ldots, 0,
2, \ldots, 2)^T$, with ten repeated values in each block of coefficients.
\item[Simulation Setting 4]: Same as Setting~3, but with $\beta = (3, \ldots, 3, 0, \ldots, 0)^T$
(15 coefficients with $\beta_j=3$ and 25 coefficients with $\beta_j = 0$), and
$V$ a block-diagonal matrix with $V_{ii} = 1.01$ for $i=1,\ldots,15$ and $V_{ii} = 1$
for $i > 15$. $V$ is split into four blocks ($1\leq i, j, \leq 5$, $6\leq i, j \leq 10$,
$11\leq i, j \leq 15$, and $i,j > 15$) with $V_{ij} = 1$ $(i \neq j)$ within the first three blocks,
$V_{ij} = 0$ $(i \neq j)$ in the fourth block, and $V_{ij} = 0$ $(i \neq j)$ across all blocks.
\end{description}

We simulated 50 data sets within each simulation setting. The response vector $y$ and the columns of
$X$ were mean-centered for each simulated data set, but the columns of $X$ were not scaled. The same
simulated data sets were used when fitting all models described below, effectively treating the simulated
data set as a blocking factor to reduce the variance of our estimates of comparative predictive performance.

To assess the impact of the matrix $\Sigma$ on predictive performance, we consider three different
approaches for specifying this parameter: $\Sigma = I_p$ (the traditional Bayesian elastic net),
$\Sigma = n(X^TX)^{-1}$ (the $\ell_1$-regularized $g$ prior scaled by $n$ to put the prior on the
unit information scale before scaling by $\lambda_2$, mimicking the Zellner--Siow prior), and an 
informative choice of $\Sigma$ based on our prior knowledge of the data analysis setting (in this case, 
the simulation setup).  The matrix $\Sigma$ describes prior dependence among the regression coefficients 
$\beta$. When specifying an informative choice 
for $\Sigma$, we should therefore focus on how we believe the regression coefficients should behave relative 
to each other.  Our informative choices of $\Sigma$ in the four simulation setups were constructed as follows.
\begin{description}
\item[Simulation 1 Informative Prior:] The vector of true regression coefficients
has no discernible pattern, and we have no
additional information to inform us about the relationship between the coefficients. We therefore
do not specify an informative prior for this simulation setting.
\item[Simulation 2 Informative Prior:] The vector of true regression coefficients has one large
block where all coefficients are equal. An informative prior would encourage strong pair-wise
dependence among all regression coefficients, and so we choose $\Sigma_{ij} = 0.8$ for $i\neq j$
and $\Sigma_{ii} = 1$.
\item[Simulation 3 Informative Prior:] The vector of true regression coefficients has two
blocks spread out across the vector, with equal coefficients within each block. An informative
prior would therefore encourage strong dependence within each block and across the similar blocks.
We therefore set
\[
    \Sigma = \left( \begin{array}{cccc}
        H & 0 & G & 0 \\
        0 & H & 0 & G \\
        G & 0 & H & 0 \\
        0 & G & 0 & H 
    \end{array}\right),
\]
where $H$ is a $10 \times 10$ matrix with $H_{ii} = 1$ and $H_{ij} = 0.5$ for $i \neq j$,
and $G$ is a $10 \times 10$ matrix with $G_{ij} = 0.5$ for all $i$ and $j$. 
\item[Simulation 4 Informative Prior:] The vector of true regression coefficients has two consecutive
blocks of equal coefficients, and so our informative prior uses
\[
    \Sigma = \left( \begin{array}{cc}
            H_1 & 0 \\
            0 & H_2
    \end{array}\right),
\]
where $H_1$ is a $15\times 15$ matrix, $H_2$ is a $25 \times 25$ matrix, the diagonal elements
of both matrices are 1, and the off-diagonal elements of both matrices are $0.5$.
\end{description}

Finally, we consider four hyperparameter settings to further understand the impact of the prior
on the two penalty terms as follows, with $\nu_a = 3$ and $\nu_b = 6$ in all four settings.
\begin{description}
\item[Hyperparameter Setting 1:] $\nu_1 = 1$, $\nu_2 = 1$, $L = (p + \nu_2)/2$, $R = \nu_2$. 
This is the $\ell_1$-regularized $t$ prior with similar penalty strength for the $\ell_1$- and 
$\ell_2$-norm components.
\item[Hyperparameter Setting 2:] $\nu_1 = 0.1$, $\nu_2 = 1$, $L = (p + \nu_2)/2$, $R = \nu_2$. 
This is the $\ell_1$-regularized $t$ prior with relatively stronger $\ell_1$-norm penalization.
\item[Hyperparameter Setting 3:] $\nu_1 = 1$, $\nu_2 = 1$, $L = 1$, $R = 1$, a uniform prior on
$\lambda_2/(\lambda_1 + \lambda_2)$ under the base prior.
\item[Hyperparameter Setting 4:] $\nu_1 = 1$, $\nu_2 = 1$, $L = 9$, $R = 1$, a base prior on
$\lambda_2/(\lambda_1 + \lambda_2)$ that favors $\ell_1$-norm penalization.
\end{description}

\begin{figure}[!t]
 \begin{center}
 \includegraphics[scale=0.5]{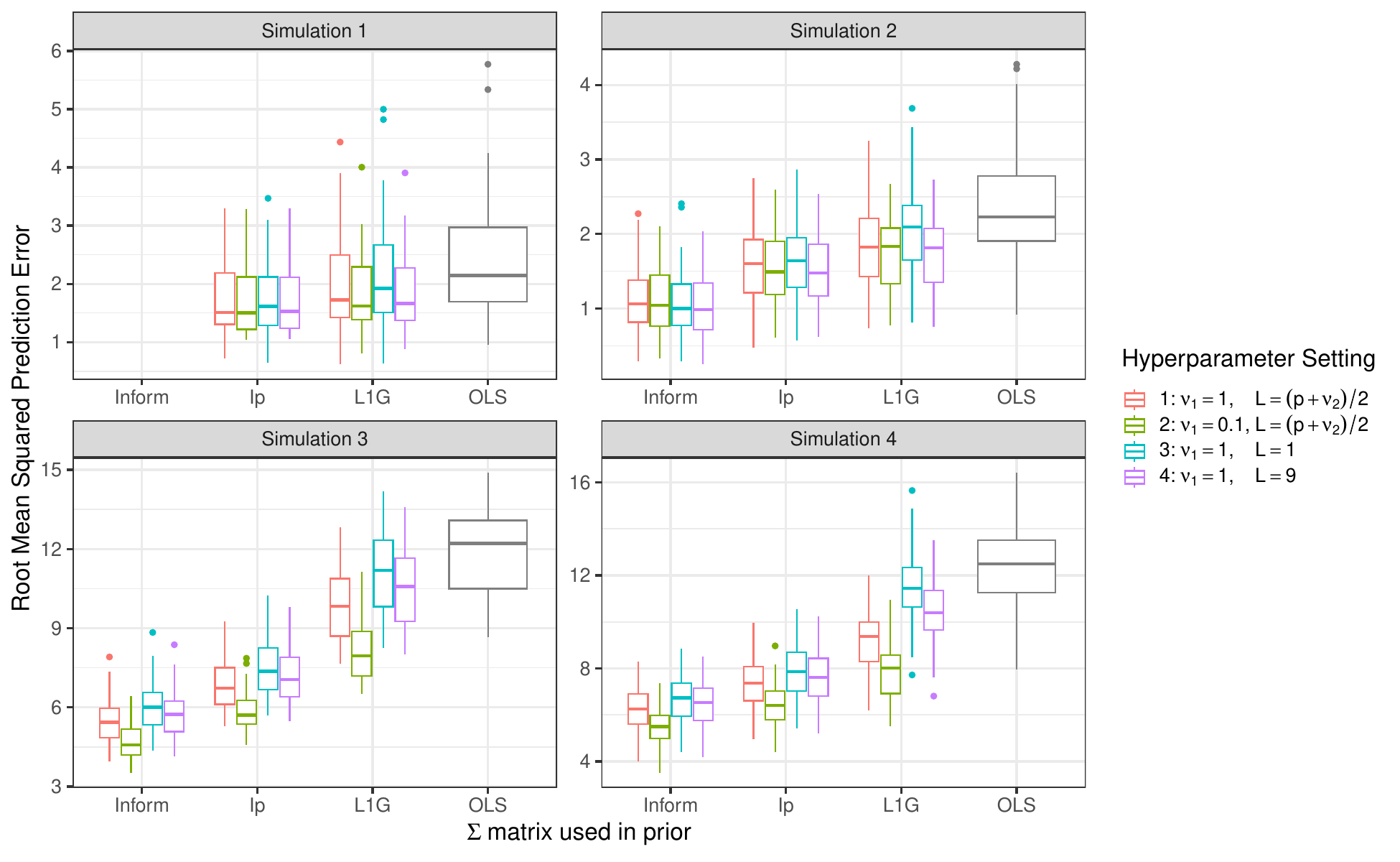}
 \caption{Root mean squared prediction error across 50 simulated data sets for the four
		  simulation settings, four hyperparameter settings, and three choices of
		  $\Sigma$ described in Section~\ref{sec:sim}. We set $R = \nu_2 = 1$
		  in each of the four hyperparameter settings. The root mean squared prediction 
		  error based on the OLS fit is shown as a reference. No informative $\Sigma$
		  matrix was specified for Simulation~1.
 	\label{fig:sim_results}}
 \end{center}
\end{figure}

Figure~\ref{fig:sim_results} shows the distribution of root mean squared prediction errors, 
$\sqrt{(\hat{\beta} - \beta)^T V (\hat{\beta} - \beta)}$, across the 50 simulated data sets
for each of the four simulation settings, $\Sigma$, and hyperparameter settings. Posterior 
means, $\hat{\beta}$, were computed based on $10,000$ MCMC samples from each posterior 
distribution. We also compute the root mean squared prediction error based on the ordinary 
least squares (OLS) fit for each simulated data set for reference.
Table~\ref{tab:relsim} summarizes the results by treating the simulated data sets as a 
blocking factor to remove nuisance variation. For each simulated data set, the percentage 
improvement in root mean squared prediction error relative to the OLS fit is computed,
and the median values across the fifty simulated data sets are reported.

In Simulation~1, a small $p$ setting where there is no discernible pattern among the true 
regression coefficients (but moderate dependence among the predictors), we find the
$\ell_1$-regularized $g$ prior performs similarly to the usual Bayesian elastic net
with $\Sigma = I_p$ (with both performing somewhat better than ordinary least squares).
We observe a similar result in Simulation~2, another small $p$ setting, but now
begin to see the benefit of incorporating structural dependence into the prior.
The informative $\Sigma$ matrix takes advantage of our knowledge that all $p=8$ 
regression coefficients should be similar, and we see a noticeable improvement in root
mean squared prediction error. The improvement is even more apparent in Simulations~3
and 4, the two higher-dimensional settings with $p=40$. The informatively-chosen
$\Sigma$ matrices tend to perform better than either the traditional Bayesian
elastic net or the $\ell_1$-regularized $g$ prior, which tends to perform least
well among the three approaches under these simulation setups.

Differences due to choice of the non-$\Sigma$ hyperparameters are less pronounced.
Adapting the amount of $\ell_1$-norm penalization to the number of predictors by
setting $L = (p + \nu_2)/2$ tends to result in strong predictive
performance, especially when $p$ is large. Strengthening the $\ell_1$-norm
penalty under the $\ell_1$-regularized $t$ prior by setting $\nu_1 = 0.1$ 
(Hyperparameter Setting 2) produced more accurate predictions in the large $p$
examples without noticeable degradation when $p$ is small.

\renewcommand{\baselinestretch}{1}
\renewcommand{\baselinestretch}{1.85}

\renewcommand{\baselinestretch}{1}
\begin{table}
	\begin{center}
\begin{tabular}{lcccccc}
  \toprule
  & \multicolumn{3}{c}{Simulation 1} & \multicolumn{3}{c}{Simulation 2} \\
  \cmidrule(lr){2-4} \cmidrule(lr){5-7} 
   &  & $I_p$ & $n(X^TX)^{-1}$ & Inform. & $I_p$ & $n(X^TX)^{-1}$ \\
  \midrule
$\nu_1=1$, $L=(p + \nu_2)/2$ & & 27.13 & 20.06 & 56.90 & 35.17 & 21.47\\
\hline
$\nu_1=0.1$, $L=(p + \nu_2)/2$ && 28.59 & 22.63 & 55.13 & 37.27 & 27.02\\
\hline
$\nu_1=1$, $L=1$ && 25.64 & 12.48 & 57.24 & 31.82 & 11.61\\
\hline
$\nu_1=1$, $L=9$ && 28.09 & 22.96 & 56.72 & 38.75 & 27.53\\
  \midrule
  & \multicolumn{3}{c}{Simulation 3} & \multicolumn{3}{c}{Simulation 4} \\
  \cmidrule(lr){2-4} \cmidrule(lr){5-7} 
   & Inform. & $I_p$ & $n(X^TX)^{-1}$ & Inform. & $I_p$ & $n(X^TX)^{-1}$ \\
  \midrule
$\nu_1=1$, $L=(p + \nu_2)/2$ & 52.67 & 41.62 & 16.41 & 49.13 & 39.83 & 22.92\\
\hline
$\nu_1=0.1$, $L=(p + \nu_2)/2$ & 60.32 & 50.62 & 31.63 & 55.56 & 47.51 & 34.71\\
\hline
$\nu_1=1$, $L=1$ & 48.69 & 35.95 & 5.75 & 44.71 & 35.18 & 7.39\\
\hline
$\nu_1=1$, $L=9$ & 51.05 & 38.84 & 10.60 & 46.88 & 37.50 & 15.49\\
  \bottomrule
\end{tabular}
	\caption{Median percent improvement in root mean squared prediction error
	compared to OLS across the fifty simulated data sets.\label{tab:relsim}}
\end{center}
\end{table}
\renewcommand{\baselinestretch}{1.85}

\section{Example: Near-infrared spectroscopy data}\label{sec:examp}
\citet{osbo:83} and \citet{osbo:84} investigated the feasibility of using 
reflectance measurements obtained via near-infrared (NIR) spectroscopy to 
predict the proportions of several components of biscuit (cookie) 
dough. The research determined that multiple linear regression models using
reflectances at a small number of selected wavelengths as predictors were useful 
for this task, and they focused on methods for selecting appropriate wavelengths.
\citet{brow:01} used Bayesian wavelet regression-based variable selection
methods to analyze the same data set, allowing for principled selection of 
bands of wavelengths. \citet{grif:11} use these data to illustrate approaches
to Bayesian variable selection based on non-convex penalty functions.

We work with the version of the data set described in 
\citet{grif:11} (though see \citet{osbo:84} and \citet{brow:01} for more detailed
descriptions of the experimental setup), which is available as the \texttt{cookie}
data set in the \texttt{R} package \texttt{ppls} \citet{krae:08, krae:25}. The data
set contains NIR reflectance measurements at $700$ wavelengths from 1100~nm to 
2498~nm for $72$ pieces of dough, as well as separate measurements of the 
compositional percentages of four components: fat, sucrose, water, and dry flour.
We focus on predicting the dry flour content using subsets of the NIR reflectance 
measurements as predictors. 

The experimental design \citep{osbo:84} involved randomization of the dough samples 
into a training data set (the first forty rows) and a test data set (the following
thirty-two rows). \citet{osbo:84} removed the samples in rows 23 of the training set
and 21 of the test set after identifying them as outliers; we do the same in our analysis. 
We therefore have $n=39$ samples in our training set and $n=31$ samples in our test
set to assess predictive accuracy. 

\begin{figure}[ht]
 \begin{center}
 \includegraphics[scale=0.4]{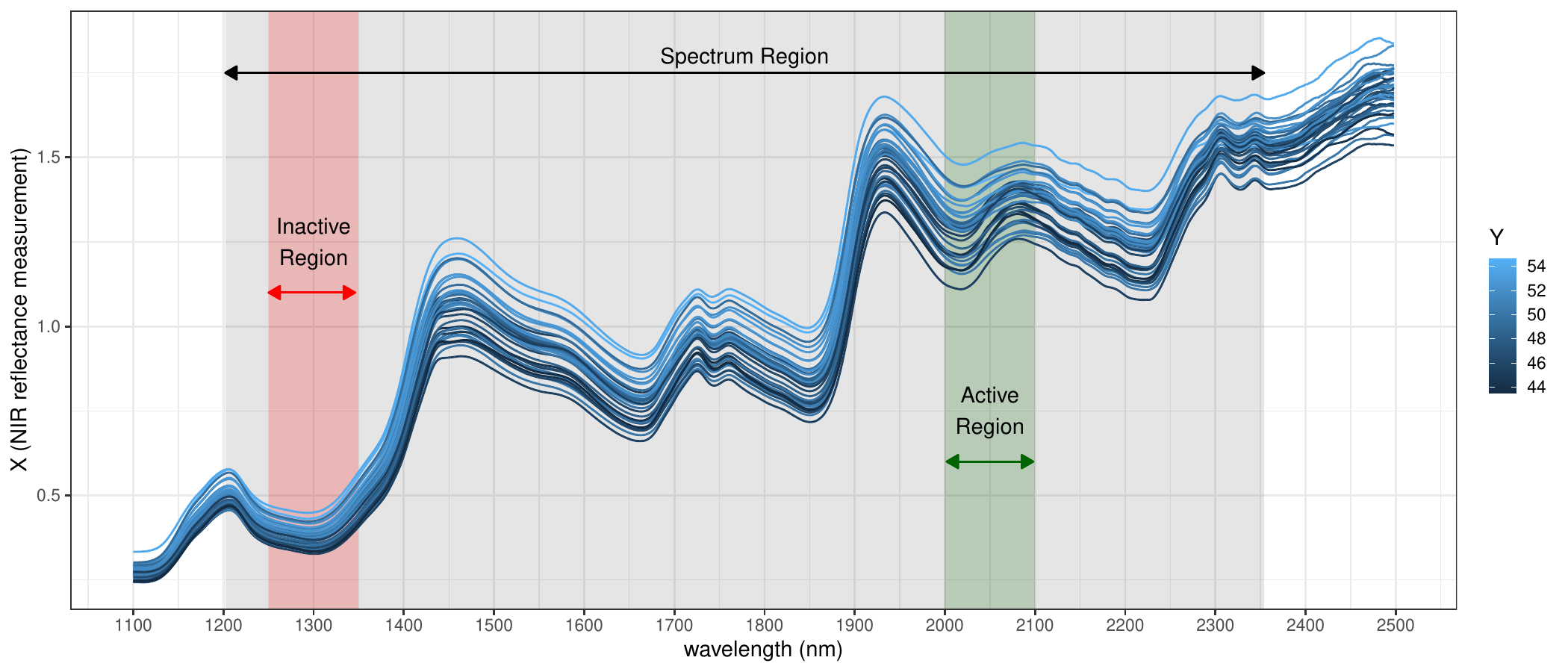}
 \caption{Each curve is one of the $n=39$ samples in the cookie dough training data
 set. The $700$ wavelengths at which the NIR reflectance measurements were taken
 are plotted on the horizontal axis, and the reflectance measurements ($X$) are plotted
 on the vertical axis. The value of the response variable, the dry flour content ($Y$),
 for each sample is indicated by the color of the curve.
 	        \label{fig:cookie_dough}}
 \end{center}
\end{figure}

The $n=39$ samples in the training data set are plotted in Figure~\ref{fig:cookie_dough}.
Each curve represents one sample, with the NIR reflectance measurement values ($X$) plotted 
on the vertical axis and the $700$ wavelengths at which the measurements were taken on the 
horizontal axis. The value of the response variable, the dry flour content ($Y$), for each 
sample is indicated by the color of its curve. 

Our aim is to demonstrate the impact on prediction accuracy of modeling prior dependence 
among regression coefficients in the Bayesian elastic net model. To this end, we consider
three different versions of the data set: one that includes samples in a narrow region of 
wavelengths that has been previously identified as being active in terms of predicting dry 
flour content, one that includes samples in a narrow region of wavelengths that has been 
previously identified as being inactive in this sense, and one that includes samples across
a large spectrum of wavelengths. \citet{grif:11} found that many of the wavelengths
identified in their analysis as being useful for predicting dry flour content were in 
the region between 1920~nm and 2080~nm. In our analysis, we take the ``active region'' 
version of the data set to contain the $p=25$ predictors measured from 2002~nm to
2098~nm, at 4~nm resolution. \citet{brow:01} excluded the region between 1100~nm and
1378~nm from their analysis, as it was ``thought to contain little useful information.''
Our ``inactive region'' version of the data set therefore contains the $p=25$ predictors
measured from 1250~nm to 1346~nm, at 4~nm resolution. The ``spectrum'' version of the data set
contains the $p=25$ predictors measured from 1202~nm to 2354~nm, at 48~nm resolution, which
is effectively the same range as used by \citet{grif:11} in their analysis (though we use a
lower wavelength resolution).  The details for the three different versions of the
data set are summarized in Table~\ref{tab:regions}, and the three wavelength regions
are indicated as the shaded regions in Figure~\ref{fig:cookie_dough}.

One might expect that, given NIR reflectances, $X$, measured at a collection
of wavelengths, the regression coefficients $\beta_j$ and $\beta_k$ might be similar
if the corresponding wavelengths, $j$ and $k$, are close to each other, and are less
likely to be similar otherwise. This suggests an AR(1)-like structure for the
dependence matrix, $\Sigma$, in the generalized orthant normal prior. For each
version of the data set, we fit the Bayesian elastic net model with
$\Sigma_{jk} = \rho^{|j-k|}$, $j, k \in \{1, ..., 25\}$, for six different values of 
$\rho \in \{0, 0.1, 0.3, 0.5, 0.7, 0.9\}$. The case $\rho = 0$ corresponds to the 
traditional Bayesian elastic net with \emph{a priori} conditional independence. Because
the active and inactive regions are sampled at 4~nm resolutions and the full spectrum 
region is sampled at a 48~nm resolution, one might consider using $\rho^{12|j-k|}$ 
as the dependence structure for the full spectrum region for comparability across
the three versions of the data set. However, the correlation would decay so rapidly
for the full spectrum data set that the results would be nearly indistinguishable 
from the independence ($\rho=0$) case. We therefore use the form for $\Sigma$ that
depends on the adjacent coefficient indices for all three versions of the data set
to allow us to explore the impact of the varying levels of prior dependence.

For all models, we specified the prior for $\sigma^2$ with $\nu_a = 4$, $\nu_b = 6$,
and used the $\ell_1$-regularized $t$ prior with $\nu_1 = 0.5$ and $\nu_2 = 2$ so
that $L = (p + \nu_2)/2 = 13.5$ and $R = \nu_2/2 = 1$. This version of the 
$\ell_1$-regularized $t$ prior places relatively more weight on the $\ell_1$-norm 
component of the penalty. The predictors, $X$, and the response $Y$ were 
mean-centered across the training data before fitting the models, but the variables 
were not scaled. Mean squared prediction errors were computed on the centered
test data set using Rao--Blackwellized posterior mean estimates of the regression 
coefficients based on $100,000$ MCMC sample after a burn-in of $5,000$ iterations.

The results are displayed in Table~\ref{tab:mspe_cookie}. For all three wavelength
regions the model with $\rho = 0.5$ produces the lowest mean squared prediction
error, demonstrating that incorporating prior dependence among the regression 
coefficients can improve predictive performance. All three wavelength regions exhibit
a similar pattern as $\rho$ varies. Incorporating even a small amount of dependence in
the model ($\rho = 0.1$) yields an improvement, with the improvement maximized at
moderate levels of dependence. It is only for very strong dependence
($\rho = 0.9$) that the predictive performance becomes worse than the conditional
independence case ($\rho = 0$). 

\renewcommand{\baselinestretch}{1}
\renewcommand{\baselinestretch}{1.85} 

\renewcommand{\baselinestretch}{1}
\begin{table}
	\begin{center}
\begin{tabular}{lcccccc}
	  \toprule
  Region & Wavelengths Included (nm) & Resolution & Number of Wavelengths ($p$) \\
  \midrule
  Active & $\{2002, 2006, ..., 2098\}$ & 4~nm & 25\\
  Inactive & $\{1250, 1254, ..., 1346\}$ & 4~nm & 25 \\
  Spectrum & $\{1202, 1250, ..., 2354\}$ & 48~nm & 25 \\
  \bottomrule
\end{tabular}
	\caption{Wavelength regions included in the three different versions of the cookie dough
	data set analyzed in Section~\ref{sec:examp}.\label{tab:regions}}
\end{center}
\end{table}
\renewcommand{\baselinestretch}{1.85}

\renewcommand{\baselinestretch}{1}
\begin{table}[!ht]
\centering
\begin{tabular}[t]{lrrrrrr}
\toprule
\multicolumn{1}{c}{ } & \multicolumn{6}{c}{Dependence Parameter} \\
\cmidrule(l{3pt}r{3pt}){2-7} 
Region & $\rho=0$ & $\rho=0.1$ & $\rho=0.3$ & $\rho=0.5$ & $\rho=0.7$ & $\rho=0.9$ \\
\midrule
Active & 5.32 & 5.26 & 5.18 & 5.12 & 5.13 & 5.35\\
Inactive & 6.11 & 6.08 & 6.02 & 5.98 & 5.99 & 6.14\\
Spectrum & 5.33 & 5.27 & 5.18 & 5.13 & 5.14 & 5.40\\
\bottomrule
\end{tabular}
\caption{Mean squared test data prediction error by wavelength region and strength of
prior dependence. The traditional Bayesian elastic net with no prior dependence
corresponds to $\rho = 0$.\label{tab:mspe_cookie}}
\end{table}
\renewcommand{\baselinestretch}{1.85}

\section{Discussion}\label{sec:disc}
This paper introduced the orthant normal distribution in its general form and showed how it can 
be used to incorporate prior dependence in the Bayesian elastic net regression model.
An $\ell_1$-regularized version of Zellner's $g$ prior was introduced as a special case, creating 
a new link between the literature on penalized optimization procedures and an important class of 
regression priors. When prior knowledge about the relationships among regression coefficients is 
available, incorporating such knowledge into the model is sound statistical practice. The new
model allows such information to be included through the dependence matrix, $\Sigma$, 

When the dependence matrix, $\Sigma$, is not a diagonal matrix, the prior distribution for
$\beta$ contains a computationally intractable normalizing constant that is a function of $\sigma^2$
and the two penalty parameters, $\lambda_1$ and $\lambda_2$. If standard priors are placed on any of
these parameters, sampling the posterior distribution becomes challenging. We resolved these
computational issues by modifying slightly the prior distribution used by \citet{hans:11}
for the Bayesian elastic net regression model. The modification ensures that the joint prior
is free of the intractable normalizing constant, allowing for straightforward and fast Gibbs
sampling. We introduced methods for assessing the impact of the modulation on the base (standard)
prior to allow the user to assess whether the new prior is appropriate for their particular analysis.
While one might object to skirting the computational issue by modifying a prior that one might have
assumed in the absence of the intractable normalizing constant, it is unlikely that the original
prior would have provided a perfect encapsulation of the user's prior beliefs. Priors for 
hyperparameters in Bayesian penalized regression models are, in most instances, priors of convenience
that have reasonable properties or provide computational advantages. Minimally modifying a prior of 
convenience to allow for simple and fast MCMC sampling seems reasonable in the context of our
model.

The simulations in Section~\ref{sec:sim} and the analysis of the NIR spectroscopy data in 
Section~\ref{sec:examp} demonstrate that incorporating prior information about dependence 
among the regression coefficients, when such information is available, can lead to improved 
predictive performance. The key, of course, is formulating an appropriate dependence structure
for a given application. In some settings, such as the NIR spectroscopy example, reasonable
dependence structures can be formulated based on one's understanding of the underlying process
being modeled. In other settings, such as the first simulation setup in Section~\ref{sec:sim},
no obvious dependence structure is apparent. If several plausible dependence structures
are available, model selection techniques could be employed to help choose among them. Such 
approaches represent an interesting avenue for future research.

\appendix
\section{Properties of the Prior Density}\label{app:densprop}
The prior density function specified by (\ref{eq:genen}) is
\begin{eqnarray*}
	p(\beta \mid \sigma^2, \lambda_1, \lambda_2) &=& 
		\sum_{z \in \mathcal{Z}} \omega_z \mbox{N}^{[z]}(\beta \mid \mu_z, \Sigma_z) \\
	&=& \sum_{z \in \mathcal{Z}} \left[\omega^{-1} \frac{\mbox{P}(z, \mu_z, \Sigma_z)}
		{\mbox{N}(0 \mid \mu_z, \Sigma_z)}\right]
		\left[ \frac{\mbox{N}(\beta \mid \mu_z, \Sigma_z)}{\mbox{P}(z, \mu_z, \Sigma_z)}
		1(\beta \in \mathcal{O}_z)\right] \\
	&=& \omega^{-1} \sum_{z \in \mathcal{Z}} \frac{\mbox{N}\left(\beta \mid -\frac{\lambda_1}{2\lambda_2}
		\Sigma z, \frac{\sigma^2}{\lambda_2}\Sigma\right)}
		{\mbox{N}\left(0 \mid -\frac{\lambda_1}{2\lambda_2}\Sigma z, \frac{\sigma^2}{\lambda_2}\Sigma\right)}
		1(\beta \in \mathcal{O}_z) \\
	&=& \omega^{-1} \sum_{z \in \mathcal{Z}} \frac{\exp\{ -\frac{\lambda_2}{2\sigma^2}
		(\beta + \frac{\lambda_1}{2\lambda_2}\Sigma z)^T \Sigma^{-1} 
		(\beta + \frac{\lambda_1}{2\lambda_2}\Sigma z)\}}
		{\exp\{-\frac{\lambda_2}{2\sigma^2}(\frac{\lambda_1}{2\lambda_2}\Sigma z)^T \Sigma^{-1} 
			(\frac{\lambda_1}{2\lambda_2}\Sigma z)\}} 1(\beta \in \mathcal{O}_z) \\
	&=& \omega^{-1} \sum_{z \in \mathcal{Z}} \frac{\exp\{ -\frac{\lambda_2}{2\sigma^2}(
		\beta^T \Sigma^{-1}\beta + \frac{\lambda_1}{\lambda_2} \beta^T z + \frac{\lambda_1^2}
			{4\lambda_2^2}z^T \Sigma z )\}}{\exp\{-\frac{\lambda_2}{2\sigma^2}
			\frac{\lambda_1^2}{4\lambda_2^2} z^T \Sigma z  \}} 1(\beta \in \mathcal{O}_z) \\
	&=& \omega^{-1} \sum_{z \in \mathcal{Z}} \exp\left\{ -\frac{\lambda_2}{2 \sigma^2}\beta^T \Sigma^{-1}
			\beta - \frac{\lambda_1}{2\sigma^2}\beta^T z \right\} 1(\beta \in \mathcal{O}_z).
\end{eqnarray*}
Noting that $\beta^T z = \sum_{j=1}^p |\beta_j| = |\beta|_1$ if $\beta \in \mathcal{O}_z$, the prior density
function is
\[
	p(\beta \mid \sigma^2, \lambda_1, \lambda_2) = \omega^{-1} \exp\left\{ - \frac{\lambda_2}{2\sigma^2}
		\beta^T \Sigma^{-1} \beta - \frac{\lambda_1}{2\sigma^2} |\beta|_1 \right\},
\]
and
\[
	-2\sigma^2 \log p(\beta \mid \sigma^2, \lambda_1, \lambda_2) = 
		\mbox{const.} + \lambda_1 | \beta |_1+ 
		\lambda_2 \beta^T \Sigma^{-1} \beta
\]
as claimed in Section~\ref{sec:penop}.  As both $\lambda_1 |\beta|_1$ and $\lambda_2 \beta^T \Sigma^{-1}
\beta$ are continuous functions of $\beta$, the prior density function $p(\beta \mid \sigma^2, \lambda_1,
\lambda_2)$ is therefore a continuous function of $\beta$. The prior density is not differentiable along the
coordinate axes when $\lambda_1 > 0$ due to the $|\beta|_1$ term.

\section{Modifications under the differentially scaled prior}\label{app:diffpar}
When the $\ell_1$- and $\ell_2$-norm penalty terms in the prior on $\beta$ are scaled differentially as in
(\ref{eq:diffpen}), the properly normalized density function is
\[
	p(\beta \mid \sigma^2, \lambda_1^*, \lambda_2) = (2\pi)^{-p/2} \left(\frac{\sigma^2}{\lambda_2}\right)^{-p/2} 
		|\Sigma|^{-1/2} \mathbb{E}\left[ e^{-\frac{\lambda^*_1}{\sqrt{\lambda_2}} |Z|_1} \right]^{-1}
		\exp\left\{ -\frac{\lambda_2}{2\sigma^2} \beta^T \Sigma^{-1} \beta - \frac{\lambda_1^*}{\sigma}
		|\beta|_1 \right\},
\]
where $Z \sim \mbox{N}\left( 0, \Sigma\right)$. Under this parameterization, we retain the prior
$p(\sigma^2) = \mbox{IG}(\sigma^2 \mid \nu_a/2, \nu_b/2)$ and modulate the base prior for $\lambda_1^*$
and $\lambda_2$ to obtain
\[
	p(\lambda_1^*, \lambda_2) \propto \mathbb{E}\left[ e^{-\frac{\lambda^*_1}{\sqrt{\lambda_2}} |Z|_1} \right]
			\times \mbox{Gamma}(\lambda_1^* \mid L, \nu_1/2) \times \mbox{Gamma}(\lambda_2 \mid R, \nu_2/2).
\]
After modulation, the density function for the joint prior distribution on $\beta$, $\sigma^2$, $\lambda_1^*$, 
and $\lambda_2$ does not depend on the term $\mathbb{E}\left[ \exp\left\{-\frac{\lambda^*_1}{\sqrt{\lambda_2}} 
|Z|_1\right\} \right]$. 

Gibbs sampling the posterior distribution under this differentially-scaled prior proceeds similarly as under
the commonly-scaled prior, but with a few modifications. When sampling $\beta_j$,
the location parameter for the positive component of the full conditional distribution changes to
\[
	\mu_j^+ = \frac{x_j^T y - (x_j^T X_{-j} + \lambda_2 \Omega_{j,-j})\beta_{-j} - \sigma \lambda_1^*}
		{\lambda_2 \Omega_{jj} + x_j^T x_j};
\]
the expression for the negative component, $\mu_j^-$, is similar but with $(-\lambda^*_1)$ replacing
$\lambda_1^*$ (c.f.~Equation~(\ref{eq:mujcomp})). The scale parameter in the full conditional distribution
for $\beta_j$ stays the same, $s^2_j = \sigma^2/(\lambda_2 \Omega_{jj} + x_j^T x_j)$.

The full conditional for $\lambda_2$ is the same under both scalings of the prior, with
\[
	\lambda_2 \mid y, \beta, \sigma^2, \lambda_1^* \sim
		\mbox{Gamma}\left( R + p/2, (\beta^T \Sigma^{-1} \beta/\sigma^2 + \nu_2)/2 \right).
\]
The full conditional for $\lambda_1^*$ under the differentially-scaled prior is slightly different from
for $\lambda_1$ under the commonly-scaled prior:
\[
	\lambda_1^* \mid y, \beta, \sigma^2, \lambda_2 \sim
		\mbox{Gamma}\left( L , |\beta|_1/\sigma + \nu_1/2 \right).
\]
Finally, the full conditional distribution for $\sigma^2$ is such that
\[
	\frac{1}{\sigma} \mid y, \beta, \lambda_1^*, \lambda_2 \sim
		\mbox{MHN}\left( p + \nu_a + n - 1, 
			\frac{(y - X\beta)^T(y - X\beta) + \lambda_2 \beta^T \Sigma^{-1}\beta + \nu_b}{2},
			\lambda_1^* |\beta|_1 \right),
\]
where $\mbox{MHN}(a,b,c)$ denotes the modified half-normal distribution for $x > 0$ with density 
$p(x) \propto x^{a-1} \exp\{-b x^2 - cx\}$, with $a > 0$, $b > 0$, and 
$c \in \mathbb{R}$. Existing algorithms allow for efficient sampling from this distribution
\citep[see, e.g., the rejection sampling methods of][]{sun:20, sun:23}.


\bibliographystyle{./natbib}
\bibliography{./genop}

\end{document}